\documentclass[12pt]{article}
\usepackage{epsf}
\usepackage{bm,physics}
\usepackage{cite}
\usepackage{amsmath,amssymb,tensor}
\usepackage{graphicx}
\usepackage[colorlinks,citecolor=blue]{hyperref}
\usepackage{comment}
\usepackage{float}
\usepackage[normalem]{ulem} %\sout
\newcommand{\m}{M_{\mathrm{Pl}}}
\newcommand{\M}{\mathrm{M}}
\newcommand{\N}{\Delta N_{\mathrm{eff}}}
\newcommand{\dif}{\mathrm{d}}

\newcommand{\MP}[1]{{\color{black}{#1}}}

\renewcommand{\thefootnote}{\fnsymbol{footnote}}
\def\thefootnote{\fnsymbol{footnote}}

\makeatletter

\@addtoreset{equation}{section}
\makeatother

\hypersetup{
    colorlinks=true,
    linkcolor=blue,      % internal links (TOC, refs)
    citecolor=blue,     % bibliography citations
    filecolor=black,   % local file links
    urlcolor=blue,       % web URLs
}

\usepackage{xcolor}
\usepackage{etoolbox}

\makeatletter
\pretocmd{\tableofcontents}
  {\hypersetup{linkcolor=black}}{}{}
\apptocmd{\tableofcontents}
  {\hypersetup{linkcolor=blue}}{}{}
\makeatother
\usepackage{titlesec}

\titleformat{\section}
  {\normalfont\large\bfseries}   % font: \large instead of default \Large
  {\thesection}{1em}{}

\titleformat{\subsection}
  {\normalfont\normalsize\bfseries}  % font: \normalsize
  {\thesubsection}{1em}{}

\allowdisplaybreaks
\begin{document}

\begin{titlepage}

\begin{center}

\vskip .45in

\bigskip\bigskip
{\Large \bf Post-Inflationary Constraints on Nonminimally Coupled Quintessential Inflation}

\vskip .65in

{\large 
Min Gi~Park,$^{1}$
Seong~Chan~Park,$^{1,2}$
Tomo~Takahashi,$^{3}$ \\ 
and Jos\'{e} Jaime Terente D\'{i}az$^{4}$
\vspace{2mm} \\
}
\vskip 0.2in

{\em 
$^{1}$Department of Physics, Yonsei University, Seoul, 03722, Republic of Korea
\vspace{2mm}\\
$^{2}$School of Physics, Korea Institute for Advanced Study, Seoul, 02455, Republic of Korea
\vspace{2mm}\\
$^{3}$Department of Physics, Saga University, Saga 840-8502, Japan
\vspace{2mm}\\
$^{4}$Faculdade de Ci\^{e}ncias e Tecnologia and CFisUC, Departamento de F\'isica, \\
Universidade de Coimbra, Rua Larga, P-3004-516 Coimbra, Portugal
}

\end{center}
\vskip .5in

\begin{abstract}
\noindent
We investigate quintessential inflation in a nonminimally coupled scalar-tensor theory, parameterizing the post-inflationary radiation abundance independently of the reheating mechanism. The nonadiabatic inflation--kination transition generates a stochastic gravitational-wave background whose contribution to $\N$ imposes a lower limit on the reheating temperature. Because this temperature dictates the duration of kination and the available scalar-field excursion, it directly constrains the present-day dark-energy equation of state. While a single-exponential coupling achieves the required post-inflationary potential drop, the same constant slope does not provide viable late-time acceleration. A double-exponential deformation resolves this tension by decoupling the average slope governing the total potential drop from the asymptotic slope driving cosmic acceleration. Full numerical solutions confirm this picture, yielding a thawing quintessence regime with $w_{\varphi,0}\simeq (-0.90, -0.95)$ for our benchmarks. Our results demonstrate that future dark-energy measurements can directly probe the post-inflationary reheating history of the Universe.
\end{abstract}
\end{titlepage}
\renewcommand{\thepage}{\arabic{page}}
\setcounter{page}{1}
\renewcommand{\thefootnote}{\#\arabic{footnote}}
\setcounter{footnote}{0}
\tableofcontents 

%%%%%%%%%%%%%%%%%%%%%%%%%%%%%%%%%%%%%%%%
\section{\label{sec:intro}Introduction}
%%%%%%%%%%%%%%%%%%%%%%%%%%%%%%%%%%%%%%%%
Cosmological observations now probe nearly the full history of cosmic evolution \cite{CosmoVerseNetwork:2025alb}. These observations point to two distinct epochs of accelerated expansion: an early phase of inflation \cite{Starobinsky:1980te,Kazanas:1980tx,Sato:1981qmu,Guth:1980zm,Linde:1981mu}, and the present era of dark-energy domination \cite{Copeland:2006wr,Frieman:2008sn,Li:2011sd,Huterer:2017buf,Garcia-Garcia:2026nzy}. Although their underlying mechanism remains elusive \cite{Dimopoulos:2020pjx}, both phenomena share a common dynamical origin in the slow evolution of a scalar field on a sufficiently flat potential \cite{Carroll:2003qq}. This striking similarity has motivated the search for unified scenarios in which a single scalar degree of freedom drives both periods of cosmic acceleration. Quintessential inflation provides one of the most compelling realizations of this framework \cite{Peebles:1998qn,deHaro:2021swo,Bettoni:2021qfs}. 

This unification, however, presents significant challenges. Specifically, the scalar potential must assume a distinct form to successfully bridge the early- and late-time accelerating eras. Such potentials are typically nonoscillatory: an inflationary plateau connects to a runaway quintessential tail, leaving no minimum around which the field can oscillate~\cite{Peloso:1999dm,Dimopoulos:2000md,Dimopoulos:2001qu,Dimopoulos:2001ix,WaliHossain:2014usl}. Consequently, reheating cannot proceed through the conventional decay of a coherently oscillating inflaton field~\cite{Allahverdi:2010xz,Lozanov:2019jxc}. This necessitates an alternative mechanism~\cite{Haro:2018jtb}, such as gravitational reheating~\cite{Ford:1986sy,Chun:2009yu,Dimopoulos:2018wfg,Haro:2018zdb,Kolb:2023ydq,Dorsch:2024nan,Jenks:2024fiu}, instant preheating~\cite{Felder:1998vq,Felder:1999pv,Campos:2002yk,Dimopoulos:2017tud}, Ricci reheating~\cite{Opferkuch:2019zbd,Bettoni:2021zhq}, curvaton reheating~\cite{Feng:2002nb,Matsuda:2007ax,BuenoSanchez:2007jxm}, reheating via primordial-black-hole evaporation~\cite{Dalianis:2021dbs,RiajulHaque:2023cqe}, or warm dynamics~\cite{Dimopoulos:2019gpz,Rosa:2019jci}. Additionally, a viable quintessential inflation model must consistently connect the inflationary energy scale to the observed dark-energy density across an energy hierarchy of more than one hundred orders of magnitude. The scalar potential must span a large, necessarily super-Planckian, field excursion \cite{Dimopoulos:2017zvq,vandeBruck:2017voa}.

A possible way to circumvent these difficulties is to embed quintessential inflation in a nonminimally coupled scalar-tensor theory \cite{Faraoni:2000wk,Wetterich:2013wza,Dimopoulos:2021xld}.\footnote{More generally, quintessential inflation has been extensively explored within modified theories of gravity \cite{vandeBruck:2017voa,Dimopoulos:2020pas,Dimopoulos:2022tvn,Dimopoulos:2022rdp,TerenteDiaz:2023iqk,TerenteDiaz:2023kgc,Dimopoulos:2025fuq,Dimopoulos:2026iwq}.} By modifying the effective geometry of field space together with the Einstein-frame scalar-field potential, nonminimal couplings to gravity can simultaneously reconcile the flat potential required for slow-roll inflation with the runaway behavior needed for late-time quintessence. After transforming to the Einstein frame, the potential develops an inflationary plateau connected to an exponential quintessential tail, allowing a single nonminimal coupling function to interpolate between the two accelerating epochs \cite{Park:2024ceu}. This construction is, moreover, of current phenomenological interest, having been confronted with recent data on the dark-energy equation of state (EoS) \cite{Wolf:2025jed,SanchezLopez:2025uzw}.

Despite these successes, several important questions remain open. In particular, the minimal single-exponential realization exhibits a tension between the requirements of reaching the observed dark-energy density and of sustaining late-time acceleration. As discussed below, this tension is controlled by the post-inflationary kination dynamics \cite{Gouttenoire:2021jhk}, which fix the scalar-field excursion connecting the inflationary and quintessential regimes. An equally important issue concerns reheating \cite{He:2018mgb,Cheong:2021kyc}. In quintessential inflation, gravitational particle production constitutes one of the most economical reheating mechanisms, requiring no additional interactions beyond gravity. However, gravitational reheating is intrinsically inefficient, since the transfer of energy into radiation is suppressed by the Planck scale \cite{deHaro:2021swo,deHaro:2023gho}. Although several alternative mechanisms have been proposed, the extent to which minimal gravitational reheating can consistently account for both the thermal history of the Universe and the late-time evolution of the scalar field remains largely unexplored. In scalar-tensor theories this question is further complicated by the conformal-frame dependence of matter couplings and particle production, although classical observables remain frame independent \cite{Magnano:1993bd,Faraoni:2006fx,Catena:2006bd,Catena:2006gk,Deruelle:2010ht,Gong:2011qe,Chiba:2013mha,Prokopec:2013zya,Postma:2014vaa,Racioppi:2021jai,Diaz:2023tma}. The equivalence at the quantum level remains under debate \cite{Shapiro:1995kt,Steinwachs:2011zs,Kamenshchik:2014waa,Falls:2018olk}.

In this work we revisit nonminimally coupled quintessential inflation from the perspective of the post-inflationary reheating history. Rather than treating reheating and late-time dark energy as independent ingredients of the cosmological evolution, we investigate how the post-inflationary kination epoch links the two. Here it is important to notice that the same nonadiabatic transition responsible for gravitational particle production also generates a stochastic gravitational-wave (GW) background \cite{Caprini:2018mtu,HaroCases:2020nsn}. Because these gravitons redshift as dark radiation, observational limits on the additional relativistic energy density, parameterized by $\N$, indirectly constrain the duration of kination. This, in turn, limits the scalar-field excursion after inflation and hence the maximum logarithmic drop before the onset of dark-energy domination. In this way, reheating and the late-time dark-energy sector become directly linked through the kination field excursion, rather than being independent parts of the model.\footnote{A related connection between the reheating history and late-time dark energy in quintessential inflation was recently explored in Ref.~\cite{An:2025ybj}, within minimally coupled models and using the approximation that the scalar field remains effectively frozen after reheating, leading to $w_{\varphi,0}\simeq -1$. In the present work, we instead study nonminimally coupled quintessential inflation, derive a lower bound on the reheating temperature from the stochastic GW contribution to $\N$, and show how this bound constrains the post-inflationary field excursion and the shape of the quintessence potential. We further construct an explicit thawing scenario in which the field unfreezes at low redshift, yielding a present-day equation of state measurably different from $-1$.} We show that this interplay has important consequences for the construction of viable quintessential inflation models, leading to nontrivial restrictions on the form of the nonminimal coupling function and the resulting late-time cosmological evolution.

Two results follow. First, the reheating history, rather than the late-time dynamics alone, determines the viability of the minimal single-exponential model. The lower bound on the reheating temperature restricts the post-inflationary field excursion. As a result, reproducing the required potential drop demands a steep average logarithmic slope, which in the single-exponential model is also the late-time slope. Consequently, the scalar field is driven toward the well-known matter-scaling attractor \cite{Copeland:1997et,SavasArapoglu:2017pyh,Bahamonde:2017ize}. Second, we show that this obstruction is removed by a minimal deformation, a double-exponential coupling function, that decouples the average slope, which fixes the integrated potential drop, from the late-time slope, which determines the present EoS. For a representative benchmark, a full numerical evolution from the end of inflation to the present epoch confirms this picture, yielding a present-day EoS \MP{$w_{\varphi,0}\simeq (-0.90, -0.95)$}, a form of evolving dark energy of the kind currently being probed by late-time surveys \cite{DESI:2025zgx}.

The paper is organized as follows. Section~\ref{sec:theoretical-framework} introduces the nonminimally coupled scalar-tensor framework, its Einstein-frame formulation, and our logarithmic-slope conventions. In Sec.~\ref{sec:postinf-reheating-general}, we analyze the post-inflationary evolution, showing how gravitational-wave bounds on $\N$ constrain the reheating temperature, kination duration, and available field excursion. Section~\ref{sec:single-exponential-model} examines a single-exponential coupling and confronts its inflationary predictions with CMB data; we show that matching the dark-energy scale forces the field onto a matter-scaling attractor, precluding late-time acceleration. In Sec.~\ref{sec:double-exponential-extension}, we introduce a double-exponential extension that decouples the integrated potential drop from the late-time equation of state, restoring viable thawing quintessence. Section~\ref{sec:numerical-evolution} confirms these analytical results via a full numerical evolution from inflation to the present epoch ($w_{\varphi,0}\simeq-0.9$). Finally, Sec.~\ref{sec:conclusions} summarizes our conclusions.

Throughout this work, we use natural units $\hbar=c=1$ and denote the reduced Planck mass by
\begin{align}
    \m
    \equiv
    (8\pi G)^{-1/2}
    \simeq
    2.44\times10^{18}\,{\mathrm{GeV}}~.
\end{align}
We adopt the mostly-plus metric signature $(-,+,+,+)$.

%%%%%%%%%%%%%%%%%%%%%%%%%%%%%%%%%%%%%%%%%%%%%%%%%%%
%%%%%%%%%%%%%%%%%%%%%%%%%%%%%%%%%%%%%%%%%%%%%%%%%%%

%%%%%%%%%%%%%%%%%%%%%%%%%%%%%%%%%%%%%%%%
\section{\label{sec:theoretical-framework}Theoretical Framework}
%%%%%%%%%%%%%%%%%%%%%%%%%%%%%%%%%%%%%%%%
\subsection{\label{subsec:jordan-einstein-frame-action}Jordan and Einstein Frames}
We consider a scalar-tensor theory of gravity in the Jordan frame defined by the following action:
\begin{align}
    S = \int \mathrm{d}^4x\,\sqrt{-g} \left[\frac{\m^2}{2}F(\phi)R -\frac{1}{2}\left(\partial\phi\right)^2
    -V(\phi) +\mathcal{L}_{I}(g_{\mu\nu},\phi,\Xi) +\mathcal{L}_{\M}(g_{\mu\nu},\Psi)\right].
    \label{eq:Jordan-frame-action-asitis}
\end{align}
Here $g_{\mu\nu}$ and $R$ denote the Jordan-frame metric tensor and Ricci scalar, respectively, and $\phi$ is the nonminimally coupled scalar field. $F(\phi)$ and $V(\phi)$ are the nonminimal coupling function and the scalar potential, respectively. The kinetic term of the scalar field is defined as
\begin{align}
    -\frac{1}{2}(\partial \phi)^2
    \equiv
    -\frac{1}{2}g^{\mu\nu}\partial_{\mu}\phi \partial_{\nu}\phi~.
\end{align}
The interaction Lagrangian $\mathcal{L}_I$ accounts for possible direct couplings between $\phi$ and a subset of the matter sector \cite{Copeland:2021qby},
collectively denoted by $\Xi$, while $\mathcal{L}_{\M}$ describes the remaining matter fields $\Psi$, which do not couple directly to $\phi$.

It is convenient to rewrite the theory in the Einstein frame, where the gravitational sector takes the standard Einstein-Hilbert form. This is achieved through a Weyl rescaling of the metric of the form \cite{Dicke:1961gz,Faraoni:1998qx,Catena:2006gk}
\begin{align}
    \label{eq:Weyl-rescaling-metrictensor}g_{\mu\nu}(x)\rightarrow\tilde{g}_{\mu\nu}(x)=F[\phi(x)]g_{\mu\nu}(x)~,
\end{align}
under which
\begin{align}
    \sqrt{-\tilde{g}(x)}=F^2[\phi(x)]\sqrt{-g(x)}~.
\end{align}
The Einstein-frame and Jordan-frame Ricci scalars satisfy \cite{Dabrowski:2008kx}
\begin{align}
    \label{eq:Ricci-rescaling-Weyl}\tilde{R}=F^{-1}\left\{R-3\left[\Box \ln F +\frac{1}{2}\left(\partial \ln F\right)^2\right]\right\},
\end{align}
with 
\begin{align}
   \label{eq:def-box-operator}\Box \equiv g^{\mu\nu}\nabla_{\mu} \nabla_{\nu}
\end{align}
being the d'Alembertian. Upon integration by parts and neglecting a boundary term, the action \eqref{eq:Jordan-frame-action-asitis} becomes\footnote{The Jordan-frame and Einstein-frame actions are denoted differently solely to emphasize the change of field variables; they correspond to the same physical action expressed in different field variables.} 
\begin{align}
    \tilde S
    &=
    \int \dif^4x\,\sqrt{-\tilde g}\,
    \Bigg[
        \frac{\m^2}{2}\tilde R
        -
        \frac{1}{2}J^2(\phi)(\tilde\partial\phi)^2
        -
        \tilde V(\phi)
        \nonumber\\
    &\hspace{7.5cm}
        +
        \tilde{\mathcal L}_I
        \left(
            \frac{\tilde g_{\mu\nu}}{F(\phi)},
            \phi,
            \Xi
        \right)
        +
        \tilde{\mathcal L}_{\M}
        \left(
            \frac{\tilde g_{\mu\nu}}{F(\phi)},
            \Psi
        \right)
    \Bigg]~,
    \label{eq:Einstein-frame-action}
\end{align}
where
\begin{eqnarray}
   J^2(\phi) &\equiv& \frac{1}{F(\phi)}\left\{1+\dfrac{3}{2}\dfrac{\m^2\left[F_{,\phi}(\phi)\right]^2}{F(\phi)}\right\},
    \label{eq:non-canonical-function}\\
    \tilde{V}(\phi) &\equiv& \dfrac{V(\phi)}{F^2(\phi)}~,
   \label{eq:potentialU}
\end{eqnarray}
and 
\begin{align}
    \label{eq:notation-tilde-square-kinetic}(\tilde{\partial} \phi)^2
    \equiv
    \tilde{g}^{\mu\nu}\partial_{\mu}\phi \partial_{\nu}\phi~.
\end{align}
A subscript comma denotes differentiation with respect to the argument; \textit{e.g.}, $F_{,\phi}(\phi)\equiv\dif F/\dif\phi$. The matter
Lagrangians transform according to
\begin{align}
    \tilde{\mathcal L}_I
    &=
    F^{-2}\mathcal L_I~,
    \\
    \tilde{\mathcal L}_\M
    &=
    F^{-2}\mathcal L_\M~.
\end{align}

To obtain a canonical kinetic term, we redefine the scalar field through
\begin{align}
    J(\phi)
    :=
    \frac{\mathrm{d}\varphi}{\mathrm{d}\phi}~.
    \label{eq:field-redefinition-canonical}
\end{align}
The Einstein-frame field $\varphi$ is therefore canonically normalized and is assumed to be an invertible function of $\phi$, although closed-form solutions to Eq.~\eqref{eq:field-redefinition-canonical} are generally difficult to obtain \cite{Mimoso:1994wn,Garcia-Bellido:2008ycs,Domcke:2017rzu,Tang:2021lcn}. The action $\tilde{S}$ can then be written as
\begin{align}
    \tilde S
    &=
    \int \dif^4x\,\sqrt{-\tilde g}\,
    \Bigg\{
        \frac{\m^2}{2}\tilde R
        -
        \frac{1}{2}(\tilde\partial\varphi)^2
        -
        \tilde V[\phi(\varphi)]
        \nonumber\\
    &\hspace{5.7cm}
        +
        \tilde{\mathcal L}_I
        \left(
            \frac{\tilde g_{\mu\nu}}
                 {F[\phi(\varphi)]},
            \phi(\varphi),
            \Xi
        \right)
        +
        \tilde{\mathcal L}_\M
        \left(
            \frac{\tilde g_{\mu\nu}}
                 {F[\phi(\varphi)]},
            \Psi
        \right)
    \Bigg\}~.
    \label{eq:EinsteinII-frame-action}
\end{align}
The modification of gravity is therefore not removed by the Weyl rescaling and the subsequent field redefinition, as it now manifests itself through the universal coupling of the matter sector to $\varphi$.  

The class of quintessential inflation models considered in this work is specified by the choice \cite{Park:2024ceu}
\begin{align}
    F(\phi)
    \equiv
    1+K(\phi)~,
    \qquad
    V(\phi)
    \equiv
    V_0K^2(\phi)~,
\end{align}
which leads to the Einstein-frame potential
\begin{align}
    \tilde V(\varphi)
    =
    V_0
    \frac{K^2[\phi(\varphi)]}
         {(1+K[\phi(\varphi)])^2}~.
\end{align}
This form naturally interpolates between two asymptotic regimes. For large values of the nonminimal coupling, $|K|\gg1$, the potential approaches an inflationary plateau, whereas in the opposite limit, $|K|\ll1$, it develops a runaway tail capable of supporting late-time quintessence for a suitable choice of $K(\phi)$:
\begin{align}
    \label{eq:thetildepotential-asymp}
    \tilde{V}(\varphi)=
    \begin{cases}
    V_0 \left[1-2/K+ {\mathcal O}(1/K^2)\right]~,
    & |K|\gg 1~,\\[8pt]
    V_0 K^2 \left[1-2K +{\mathcal O}(K^2)\right]\ll V_0~,
    & |K|\ll 1~.
    \end{cases}
\end{align}
The specific form of $K(\phi)$ determines the detailed inflationary and late-time dynamics, and is introduced in Secs.~\ref{sec:single-exponential-model} and \ref{sec:double-exponential-extension}.

%%%%%%%%%%%%%%% 
%%%%%%%%%%%%%%%

\subsection{\label{subsec:logarithmic-slope-potential}Logarithmic Slope and Field Convention}
To characterize the Einstein-frame potential, it is convenient to introduce its logarithmic slope with respect to the canonically normalized field:
\begin{align}
    \lambda_\varphi\equiv-\frac{\m}{\tilde V}\frac{\dif\tilde V}{\dif\varphi}~.
    \label{eq:lambda-varphi}
\end{align}
Since the model is originally specified in terms of the Jordan-frame field $\phi$, we also define the logarithmic slope with respect to the original field coordinate:
\begin{align}
    \lambda_\phi\equiv-\frac{\m}{\tilde V}\frac{\dif\tilde V}{\dif\phi}~.
    \label{eq:lambdaphi}
\end{align}
Using the field redefinition in Eq.~\eqref{eq:field-redefinition-canonical}, the two slopes are related through
\begin{align}
    \lambda_\varphi
    =\frac{\lambda_\phi}{J(\phi)}~.
    \label{eq:slope-field-relation}
\end{align}
Consequently,
\begin{align}
    \lambda_\varphi\,\dif\varphi=\lambda_\phi\,\dif\phi~,
    \label{eq:slope-measure-invariance}
\end{align}
showing that the logarithmic potential drop is invariant under field reparameterizations.

During inflation, the distinction between the Jordan-frame field $\phi$ and the canonically normalized Einstein-frame field $\varphi$ is essential, since the field-space factor $J(\phi)$ generally differs significantly from unity. In the present scenario, however, inflation is followed by a kination phase (see Sec.~\ref{sec:kination-reheating-spec}). By the onset of kination, the numerical evolution yields (the subscript ``$\mathrm{kin}$'' denotes evaluation at the onset of kination)
\begin{align}
    J(\phi_{\mathrm{kin}})=\left.\frac{\dif\varphi}{\dif\phi}\right|_{\mathrm{kin}}\simeq 1~,
\end{align}
and $J(\phi)$ remains very close to unity throughout the subsequent evolution. We therefore describe the inflationary dynamics and the numerical evolution in terms of the canonical field $\varphi$. In the analytic post-inflationary treatment, where $J\simeq1$, we identify $\phi\simeq\varphi$, which considerably simplifies the analysis without affecting the leading-order dynamics.

%%%%%%%%%%%%%%%%%%%%%%%%%%%%%%%%%%%%%%%%%%%%%%%%%%%
%%%%%%%%%%%%%%%%%%%%%%%%%%%%%%%%%%%%%%%%%%%%%%%%%%%

%%%%%%%%%%%%%%%%%%%%%%%%%%%%%%%%%%%%%%%%
\section{\label{sec:postinf-reheating-general}Post-Inflationary Evolution and the Reheating Bound}
As discussed in Sec.~\ref{subsec:logarithmic-slope-potential}, the field-space factor satisfies $J(\phi_{\mathrm{kin}})\simeq1$ at the onset of kination. This is the post-inflationary regime where $|K|\ll1$, so that $F\simeq1$ and $F_{,\phi}\to0$. We therefore identify $\phi\simeq\varphi$ throughout the following analytic treatment of the post-inflationary evolution. The covariant field equations and the corresponding background equations in the two frames are given in Appendix~\ref{app:cov-field-eqs-background}. Throughout this section, we use
\begin{align}
    N\equiv\ln(a/a_{\mathrm{kin}})
\end{align}
as the time variable, so that $N=0$ at the onset of kination.
Since $F\simeq1$ in this regime, the distinction between the
Jordan- and Einstein-frame $e$-fold numbers is negligible at the order
considered here.
%%%%%%%%%%%%%%%%%%%%%%%%%%%%%%%%%%%%%%%%
\subsection{\label{sec:kination-reheating-spec}Kination and Reheating}
After the end of inflation, the Universe enters a kination phase, during which the kinetic energy of the scalar field dominates over its potential energy, $\tilde V\ll\dot\phi^2/2$. The scalar field therefore behaves as a stiff fluid with EoS parameter
\begin{align}
    w_\phi\simeq 1~,
\end{align}
so that its energy density redshifts as
\begin{align}
    \rho_\phi(N)\simeq\rho_{\mathrm{kin}}(N)=\rho_{\phi,\mathrm{kin}}e^{-6N}~.
\end{align}
This scaling follows directly from the Klein-Gordon equation whenever the potential-gradient term is negligible, and is independent of whether the scalar field dominates the total energy density.\footnote{Whenever \(\ddot{\phi}+3H\dot{\phi}\simeq0\,,\) one has \(\dot{\phi}\propto a^{-3}\,,\) and hence \(\rho_{\mathrm{kin}}=\dot{\phi}^2/2\propto a^{-6}\,,\) independently of the dominant background component.}

The scalar-field excursion per $e$-fold is determined by the Friedmann equation:
\begin{align}
    \left.
    \begin{aligned}
        H^2
        &=
        \frac{\rho_{\mathrm{tot}}}{3\m^2}
        \\
        \rho_{\mathrm{kin}}
        &=
        \frac{1}{2}\dot{\phi}^2
    \end{aligned}
    \right\}
    \quad\Longrightarrow\quad
    \phi'
    \equiv
    \frac{\dif\phi}{\dif N}
    =
    \frac{\dot{\phi}}{H}
    =
    \sqrt{6}\,\m
    \left(
        \frac{\rho_{\mathrm{kin}}}
             {\rho_{\mathrm{tot}}}
    \right)^{1/2}
    \simeq
    \sqrt{6}\,\m~,
    \label{eq:phiprime}
\end{align}
where the last approximation follows from
$\rho_{\mathrm{kin}}\simeq\rho_{\mathrm{tot}}$ during kination. Thus, the duration of kination directly determines the post-inflationary field displacement. We assume that a subdominant radiation component is already present at the onset of kination, without specifying either its production time or the underlying reheating mechanism. Such radiation may originate from gravitational particle production, instant preheating, or the production of superheavy particles followed by their decay into relativistic species~\cite{deHaro:2021swo}. Since the radiation component remains subdominant throughout kination, its detailed production history has a negligible impact on the subsequent background evolution. The efficiency of reheating is therefore fully characterized by its initial abundance,
\begin{align}
    \Theta\equiv\frac{\rho_{r,\mathrm{kin}}}{\rho_{\phi,\mathrm{kin}}}~,
    \label{eq:Theta}
\end{align}
which we treat as a free parameter satisfying
\begin{align}
    \Theta\ll 1~.
\end{align}

The scalar-field and radiation energy densities evolve as
\begin{align}
    \rho_\phi(N)
    &=
    \rho_{\phi,\mathrm{kin}}e^{-6N}~,
    \\[8pt]
    \rho_r(N)
    &=
    \rho_{r,\mathrm{kin}}e^{-4N}~.
\end{align}
We define the end of kination and the reheating epoch by
radiation--kination equality:
\begin{align}
    \rho_r(N_{\mathrm{RH}})=\rho_\phi(N_{\mathrm{RH}})~,
\end{align}
yielding
\begin{align}
    N_{\mathrm{RH}}=\frac{1}{2}\ln\left(\frac{\rho_{\phi,\mathrm{kin}}}{\rho_{r,\mathrm{kin}}}\right)=-\frac12\ln\Theta~.
    \label{eq:nkin-analytic}
\end{align}
The duration of kination therefore depends solely on the initial radiation fraction. Combining Eqs.~\eqref{eq:phiprime} and \eqref{eq:nkin-analytic}, the corresponding field excursion is
\begin{align}
\frac{\Delta\phi_{\mathrm{kin}}}{\m}\simeq\sqrt6\,N_{\mathrm{RH}}=-\frac{\sqrt6}{2}\ln\Theta~,
\label{eq:f-excursion-NRH-sqrt6}
\end{align}
which sets the scale of the post-inflationary field displacement.

We define the reheating temperature $T_{\mathrm{RH}}$ as the temperature of the radiation bath at radiation--kination equality, which we take as the onset of radiation domination. At this epoch,
\begin{align}
    \rho_{\mathrm{RH}}\equiv \rho_r(N_{\mathrm{RH}})=\rho_\phi(N_{\mathrm{RH}})=\rho_{r,\mathrm{kin}}e^{-4N_{\mathrm{RH}}}=\rho_{\phi,\mathrm{kin}}\Theta^3~.
\end{align}
Assuming that the radiation bath is in thermal equilibrium,
\begin{align}
    \rho_{\mathrm{RH}}=\frac{\pi^2}{30}g_*(T_{\mathrm{RH}})T_{\mathrm{RH}}^4~,
\end{align}
where $g_*(T)$ denotes the effective number of relativistic degrees of freedom contributing to the energy density, we obtain
\begin{align}
    T_{\mathrm{RH}}=\left[\frac{30\rho_{\phi,\mathrm{kin}}\Theta^3}{\pi^2g_*(T_{\mathrm{RH}})}\right]^{1/4}~.
    \label{eq:Tre}
\end{align}
For the high reheating temperatures considered in the numerical
analysis, we take $g_*(T_{\mathrm{RH}})=106.75$.
Given $\rho_{\phi,\mathrm{kin}}$, the reheating temperature is then set by the initial radiation fraction:
\begin{align}
    T_{\mathrm{RH}}=\left(\frac{30\,\rho_{\phi,\mathrm{kin}}}{\pi^2\times106.75}\right)^{1/4}\Theta^{3/4}~,
\end{align}
so that $T_{\mathrm{RH}}$ and $\Theta$ provide equivalent parameterizations of the initial radiation abundance.

Finally, the reheating temperature determines the Einstein-frame number of $e$-folds, $\tilde{N}_\star$, between the horizon exit of the CMB pivot scale and the end of inflation. For a reheating phase characterized by an effective EoS parameter, $w_{\mathrm{RH}}$, one finds~\cite{Cook:2015vqa} (see also Ref.~\cite{Cheong:2021kyc} for nonminimally coupled models) 
\begin{align}
    \tilde{N}_\star=61.4+\frac{3w_{\mathrm{RH}}-1}{12(1+w_{\mathrm{RH}})}\ln\left[\frac{45\tilde{V}_{\mathrm{end}}}{\pi^2g_*(T_{\mathrm{RH}})T_{\mathrm{RH}}^4}\right]-\ln\left(\frac{\tilde{V}_{\mathrm{end}}^{1/4}}{H_\star}\right).
    \label{eq:Ne}
\end{align}
In the kination-dominated regime, $w_{\mathrm{RH}}=1$. Consequently, specifying the reheating temperature not only fixes the inflationary prediction in the $n_s$--$\,r$ plane through $\tilde{N}_\star$, but also determines the duration of kination and the available post-inflationary field excursion.

%%%%%%%%%%%%%%% 
%%%%%%%%%%%%%%%

\subsection{Gravitational Particle Production and Gravitational-Wave Generation}
Gravitational particle production \cite{Kolb:2023ydq} relies on the nonadiabatic evolution of the background spacetime during the transition from inflation to kination. Vacuum fluctuations are then converted into real particles, generating a relativistic component whose energy density provides the initial radiation bath after inflation. In this section we set up the computation of the efficiency of this process and determine the corresponding radiation abundance at the onset of kination.

We begin by considering a massless minimally coupled scalar field $\chi$. Introducing the canonically normalized Fourier mode
\begin{align}
    u_k\equiv a\,\chi_k~,
\end{align}
its evolution in conformal time $\eta$ is governed by
\begin{align}
    u_k^{\prime\prime}+\omega_k^2(\eta)u_k=0~, \ \ \mathrm{with} \ \ \omega_k^2 \equiv k^2-\frac{a^{\prime\prime}}{a}~.
    \label{eq:minimally-coupled-mode-equation}
\end{align}
In this subsection, a prime denotes differentiation with respect to conformal time. The nonadiabatic transition mixes positive- and negative-frequency solutions. After the transition, each mode can therefore be written as
\begin{align}
    u_k(\eta) = \alpha_k u_k^{(+)}(\eta)+\beta_k u_k^{(-)}(\eta)~,
    \label{eq:bogoliubov-decomposition}
\end{align}
where the Bogoliubov coefficients satisfy
\begin{align}
    |\alpha_k|^2-|\beta_k|^2=1~,
\end{align}
and the occupation number of the produced particles is
\begin{align}
    n_k=|\beta_k|^2~.
\end{align}
Once the background returns to an adiabatic regime, particle production ceases and the produced particles behave as free radiation with energy density
\begin{align}
    \rho_\chi=\frac{1}{2\pi^2a^4}\int_0^\infty \dif k\,k^3|\beta_k|^2~.
    \label{eq:produced-scalar-energy-density}
\end{align}
It is convenient to absorb the details of the production spectrum into the dimensionless coefficient
\begin{align}
    C\equiv\frac{1}{2\pi^2}\int_0^\infty \dif\ln x\,x^4|\beta_k|^2~, \ \ \mathrm{with} \ \ x
    \equiv
    \frac{k}{a_{\mathrm{end}}H_{\mathrm{end}}}~,
    \label{eq:Cend-definition}
\end{align}
which measures the efficiency of gravitational particle production. The corresponding energy density at the end of inflation is simply
\begin{align}
    \rho_{\chi,\mathrm{end}}=CH_{\mathrm{end}}^4~.
    \label{eq:rho-produced-Cend}
\end{align}

Strictly speaking, particle production is completed only after the background has evolved into the adiabatic kination regime. Since the produced radiation remains completely negligible throughout this interval, its backreaction on the expansion may be ignored. It is therefore convenient to redshift the produced energy density to the onset of kination and use it to define the initial radiation abundance:
\begin{align}
    \rho_{\chi,\mathrm{kin}}=\rho_{\chi,\mathrm{end}}
    e^{-4(0-N_{\mathrm{end}})}=CH_{\mathrm{end}}^4e^{4N_{\mathrm{end}}}~,
    \label{eq:rho-produced-kination-onset}
\end{align}
where $N_{\mathrm{end}}<0$ denotes the $e$-fold coordinate at the end of inflation, whereas $N=0$ corresponds to the onset of kination, as indicated at the beginning of Sec.~\ref{sec:postinf-reheating-general}. The above calculation applies directly to primordial tensor perturbations. Writing the perturbed metric as
\begin{align}
    \dif s^2=a^2(\eta)\left\{-\dif\eta^2+\left[\delta_{ij}+h_{ij}(\eta)\right]\dif x^i \dif x^j\right\},
\end{align}
with 
\begin{align}
    \partial_i h_{ij}=0~, \ \ \mathrm{and} \ \ h_{ii}=0~,
\end{align}
each tensor polarization satisfies 
\begin{align}
    v_{\lambda,k}^{\prime\prime}+\left(k^2-\frac{a^{\prime\prime}}{a}\right)v_{\lambda,k}=0~,
    \label{eq:tensor-mode-equation}
\end{align}
where $v_{\lambda,k} \equiv \frac{a\m}{2}h_{\lambda,k}$, and $\lambda=+,\times$ indicates the two tensor polarizations \cite{Baumann:2009ds}. Equation \eqref{eq:tensor-mode-equation} is identical to Eq.~\eqref{eq:minimally-coupled-mode-equation}, and the GW energy density follows immediately from the scalar-field calculation \cite{Hashiba:2018iff,deHaro:2019oki}. Including both tensor polarizations gives
\begin{align}
    \rho_{\mathrm{GW}}=2\rho_{\chi}^{\mathrm{1dof}}~.
    \label{eq:gw-two-polarizations}
\end{align}
In particular, at the onset of kination:
\begin{align}
    \rho_{\mathrm{GW},\mathrm{kin}}=2CH_{\mathrm{end}}^4e^{4N_{\mathrm{end}}}~.
    \label{eq:rho-gw-kination-onset}
\end{align}

\begin{figure}[t]
\centering
    \includegraphics[width=\linewidth]{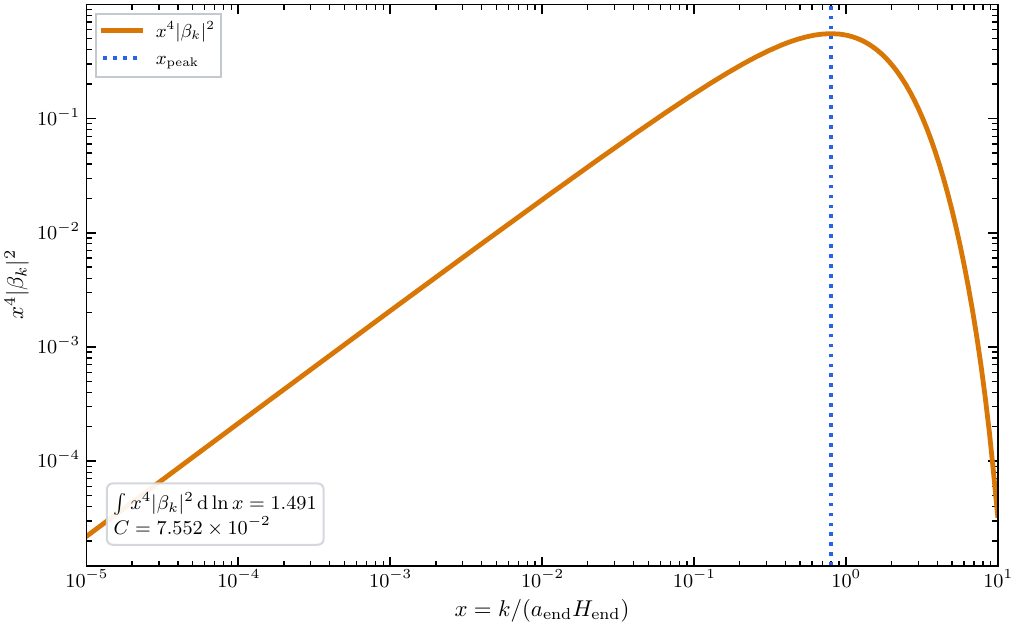}
    \caption{
        Dimensionless gravitational particle-production spectrum for
        the representative value $f_0=0.3\,\m$.
        The horizontal variable is
        $x=k/(a_{\mathrm{end}}H_{\mathrm{end}})$, and the plotted quantity is
        $x^4|\beta_k|^2$ for one massless minimally coupled degree of
        freedom.
        The coefficient $C$ is proportional to the area
        under this curve when integrated over $\dif \ln x$.
        For GWs, the total produced energy density
        includes a factor of two from the two tensor polarizations.
    }
    \label{fig:gw_production_f03}
\end{figure}

\begin{figure}[t]
    \centering
    \includegraphics[width=\linewidth]{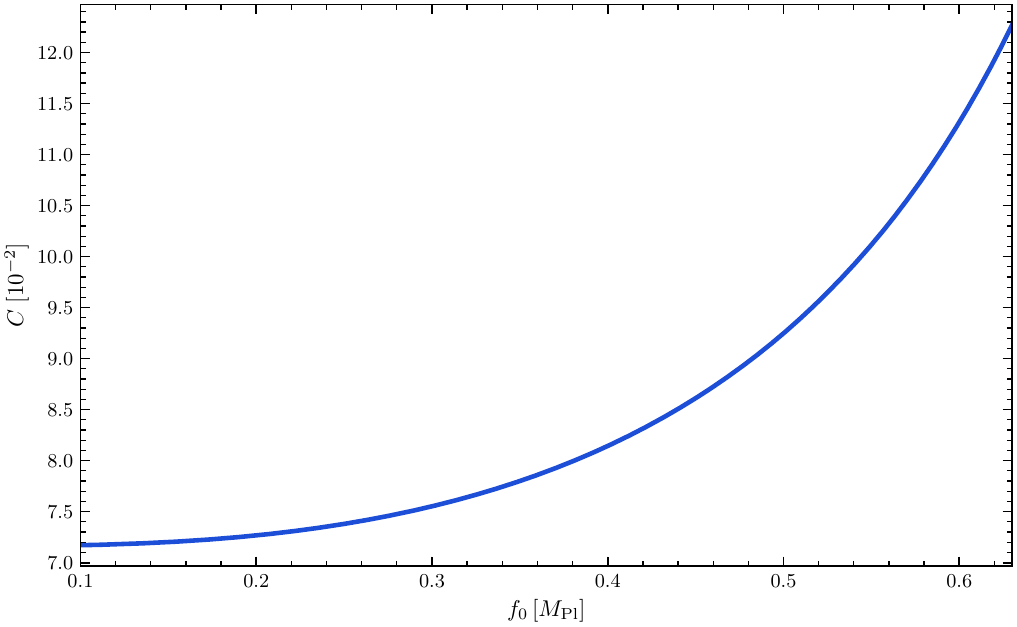}
    \caption{
        Dimensionless particle-production coefficient $C$ as
        a function of $f_0$ in Planck units.
        For each value of $f_0$, the Bogoliubov coefficient is obtained
        by solving the mode equation across the inflation--kination
        transition, and $C$ is evaluated from
        \(
        C
        =
        (2\pi^2)^{-1}
        \int \dif \ln x\,x^4|\beta_k|^2
        \).
        The plotted values correspond to one massless minimally coupled
        degree of freedom. The GW energy density is
        obtained by multiplying the result by two to account for the
        two tensor polarizations.
    }
    \label{fig:C_end_values}
\end{figure}

The formalism above is model independent, whereas the numerical value of $C$ depends on the detailed background evolution across the inflation--kination transition. For illustration, we evaluate the production spectrum using the single-exponential model introduced in Sec.~\ref{sec:single-exponential-model}, with the representative value $f_0=0.3\,\m$ (see Eq.~\eqref{eq:K_single_exp} for the exponential coupling $K(\phi)$). Figure \ref{fig:gw_production_f03} shows the resulting dimensionless spectrum $x^4|\beta_k|^2$. Since $C$ is proportional to the area under this curve when plotted against $\ln x$, repeating the calculation for different values of $f_0$ yields the function $C(f_0)$ shown in Fig.~\ref{fig:C_end_values}.

In the next subsection we combine this result with the background evolution derived previously to determine the reheating temperature and the associated contribution to $\N$.

%%%%%%%%%%%%%%% 
%%%%%%%%%%%%%%%

\subsection{The Reheating Bound on \texorpdfstring{$T_{\mathrm{RH}}$}{TRH}}
The stochastic GW background generated during the inflation--kination transition constitutes an irreducible dark-radiation component. Its energy density is therefore constrained by observational limits on the additional relativistic species, parameterized by $\N$. As we show below, this constraint places a lower bound on the initial radiation abundance and consequently on the reheating temperature, thereby limiting the duration of kination and the post-inflationary field excursion.

The total radiation density after electron--positron annihilation can be written in terms of the effective number of relativistic species,
$N_{\mathrm{eff}}$,\footnote{Not to be confused with the number of $e$-folds in the Jordan frame, $N$, used in this work.}
\begin{align}
    \rho_r=\rho_\gamma\left[1+c_\nu N_{\mathrm{eff}}\right], \ \ \mathrm{with} \ \ c_\nu\equiv\frac{7}{8}\left(\frac{4}{11}\right)^{4/3}\simeq0.2271~.
\end{align}
An additional GW background contributes as dark radiation,
\begin{align}
    \rho_{\mathrm{GW}}=c_\nu\N\,\rho_\gamma~,
\end{align}
so that
\begin{align}
    \left.\frac{\rho_{\mathrm{GW}}}{\rho_r}\right|_{\mathrm{late}}=\frac{c_\nu\N}{1+c_\nu N_{\mathrm{eff}}^{\mathrm{SM}}}\simeq0.134\,\N~,
    \label{eq:rhoGW-rhor-DeltaNeff}
\end{align}
where $N_{\mathrm{eff}}^{\mathrm{SM}}=3.044$. Current BBN and CMB observations constrain any additional relativistic component to $\N\lesssim0.3$ at $95\%$~CL \cite{Planck:2018vyg,Schoneberg:2024ifp}, while forthcoming surveys such as CMB-S4 are expected to reach $\sigma(\N)\simeq0.06$ \cite{CMB-S4:2016ple}. In what follows we adopt the fiducial bound
\begin{align}
    \N\lesssim0.1~,
\end{align}
which is conservative with respect to present data and within reach of forthcoming surveys \cite{Yeh:2022heq,Planck:2018vyg,Pitrou:2018cgg}. Taking the representative value $\Delta N_{\mathrm{eff}}^{\mathrm{max}}=0.1$ gives
\begin{align}
    \left.\frac{\rho_{\mathrm{GW}}}{\rho_r}\right|_{\mathrm{late}}\lesssim 1.3\times10^{-2}~.
    \label{eq:gw-late-observational-bound}
\end{align}
During kination both the GW background and the ordinary radiation bath redshift as $a^{-4}$, so that their ratio remains constant until the onset of radiation domination,
\begin{align}
    \left.\frac{\rho_{\mathrm{GW}}}{\rho_r}\right|_{\mathrm{kin}}=\left.\frac{\rho_{\mathrm{GW}}}{\rho_r}\right|_{\mathrm{RH}}~.
\end{align}
Subsequently, entropy conservation modifies this ratio through the change in the effective number of relativistic degrees of freedom:
\begin{align}
    \left.\frac{\rho_{\mathrm{GW}}}{\rho_r}\right|_{\mathrm{late}}=\left.\frac{\rho_{\mathrm{GW}}}{\rho_r}\right|_{\mathrm{kin}}\frac{g_{*s}^{4/3}(T_{\mathrm{late}})/g_*(T_{\mathrm{late}})}{g_{*s}^{4/3}(T_{\mathrm{RH}})/g_*(T_{\mathrm{RH}})}~.
    \label{eq:gw-radiation-gstar}
\end{align}
Here $g_{*s}$ is the analogous effective number for the entropy density; it coincides with $g_*$ at $T_{\mathrm{RH}}$ (when all relativistic species share a common temperature), but differs at late times: after $e^\pm$ annihilation, the neutrinos are colder than the photons, $T_\nu=(4/11)^{1/3}\,T_\gamma$ (see Ref.~\cite{Baumann:2018muz}), and the energy and entropy densities weight them differently. Taking \cite{Husdal:2016haj}
\begin{align}
    g_{*s}(T_{\mathrm{RH}})=g_*(T_{\mathrm{RH}})\simeq106.75~, \ \ g_{*s}(T_{\mathrm{late}})\simeq3.91~, \ \ \mathrm{and} \ \ g_*(T_{\mathrm{late}})\simeq3.36~,
\end{align}
the observational limit translates into the requirement
\begin{align}
    \left.\frac{\rho_{\mathrm{GW}}}{\rho_r}\right|_{\mathrm{kin}}\lesssim3.5\times10^{-2}~.
\end{align}
The precise numerical value depends weakly on the reference epoch adopted, but remains at the level of a few times $10^{-2}$. Throughout this work we adopt the more conservative fiducial condition
\begin{align}
    \left.\frac{\rho_{\mathrm{GW}}}{\rho_r}\right|_{\mathrm{kin}}\leq10^{-2}~,
    \label{eq:gw-radiation-initial-bound}
\end{align}
or equivalently,
\begin{align}
    \rho_{r,\mathrm{kin}}\geq100\,\rho_{{\mathrm{GW}},\mathrm{kin}}~.
    \label{eq:radiation-gw-initial-condition}
\end{align}
This condition is sufficient to satisfy the observational bound on $\N$ once the subsequent change in the number of relativistic degrees of freedom is taken into account, so that our bound on $T_{\mathrm{RH}}$ is conservative and remains valid under future tightening of the $\N$ constraint. Using Eq.~\eqref{eq:Theta}, the condition above gives
\begin{align}
    \Theta\geq100\frac{\rho_{{\mathrm{GW}},\mathrm{kin}}}{\rho_{\phi,\mathrm{kin}}}\equiv\Theta_{\mathrm{GW}}~.
    \label{eq:theta-gw-bound}
\end{align}

Equation \eqref{eq:theta-gw-bound} constitutes the first key result of our analysis. The stochastic GW background fixes a minimum radiation abundance at the onset of kination, independently of the detailed reheating mechanism. Since the reheating temperature is uniquely determined by the initial radiation fraction through Eq.~\eqref{eq:Tre}, the lower bound on $\Theta$ translates directly into one on $T_{\mathrm{RH}}$: 
\begin{align}
    T_{\mathrm{RH}}\geq\left[\frac{30\rho_{\phi,\mathrm{kin}}}{\pi^2g_*(T_{\mathrm{RH}})}\right]^{1/4}\Theta_{\mathrm{GW}}^{3/4}\equiv T_{\mathrm{RH}}^{\mathrm{GW}}~.
    \label{eq:gw-lower-bound-TRH}
\end{align}
The physical origin of this bound is straightforward: the GW abundance is fixed by the nonadiabatic inflation--kination transition, whereas decreasing the reheating temperature corresponds to reducing the initial radiation abundance. If reheating is too inefficient, GWs constitute an unacceptably large fraction of the total radiation density, violating the observational bound on $\N$.

Crucially, the implications extend well beyond reheating itself. Since the initial radiation abundance determines the duration of kination through Eq.~\eqref{eq:nkin-analytic}, the GW bound implies
\begin{align}
    N_{\mathrm{RH}}\leq-\frac{1}{2}\ln\Theta_{\mathrm{GW}}~,
    \label{eq:gw-upper-bound-Nkin}
\end{align}
placing an upper bound on the duration of the kination epoch and, via Eq.~\eqref{eq:f-excursion-NRH-sqrt6}, on the post-inflationary scalar-field excursion. This observation provides the central link explored in the remainder of this work. The observational bound on dark radiation is not merely a consistency condition on reheating: it propagates through the post-inflationary evolution to constrain the available field excursion before radiation domination. Observational constraints on reheating therefore translate directly into constraints on the shape of the quintessence potential and, ultimately, on its late-time dynamics.

%%%%%%%%%%%%%%% 
%%%%%%%%%%%%%%%

\subsection{Freezing and the Required Potential Drop}
At $N=N_{\mathrm{RH}}$, the radiation and scalar kinetic energy densities become equal, marking the end of the kination era and the onset of radiation domination. Thereafter the radiation density rapidly becomes the dominant component, while the scalar kinetic energy continues to redshift as $a^{-6}$, so that $\rho_{\mathrm{kin}}\ll\rho_{\mathrm{tot}}$. Consequently, the general expression in Eq.~\eqref{eq:phiprime} gives
\begin{align}
    \phi'=\sqrt{6}\,\m \left(\frac{\rho_{\mathrm{kin}}}{\rho_{\mathrm{tot}}}\right)^{1/2}\to 0~,
\end{align}
showing that Hubble friction rapidly damps the scalar motion. The field therefore approaches an approximately constant value, $\phi_{\mathrm{freeze}}$, and remains effectively frozen throughout the radiation-dominated epoch.

Neglecting the potential gradient but retaining both the scalar
kinetic and radiation energy contributions in the Friedmann equation, Eq.~\eqref{eq:phiprime} yields 
\begin{align}
\frac{\dif \phi}{\dif N}
    =
    \frac{\sqrt{6}\,\m}
    {\sqrt{1+\Theta e^{2N}}}~,
\end{align}
from which the total field excursion from the onset of kination to freezing is
  \begin{align}
    \Delta\phi
    =
    \sqrt{6}\,\m
    \int_0^\infty
    \frac{\dif N}{\sqrt{1+\Theta e^{2N}}}
    =
    \sqrt{6}\,\m
    \sinh^{-1}\!\left(\Theta^{-1/2}\right)
    \simeq\sqrt{6}\,\m\ln \left(\frac{2}{\sqrt{\Theta}}\right),
\end{align}
where the last equality assumes $\Theta\ll1$. Using $N_{\rm RH}=-\frac{1}{2}\ln\Theta$ (see Eq.~\eqref{eq:nkin-analytic}), this becomes
\begin{align}
\Delta \phi\simeq \sqrt{6}\, \m\left(N_{\mathrm{RH}}+\ln 2\right).
\label{eq:sf_loses_kin_en}
\end{align}
Note that $\Delta\phi$ exceeds the excursion in Eq.~\eqref{eq:f-excursion-NRH-sqrt6}:
\begin{align}
    \Delta\phi\simeq\Delta\phi_{\mathrm{kin}}+\sqrt6\,\m\ln2~.
\end{align}

It is $\Delta\phi$, rather than $\Delta\phi_{\mathrm{kin}}$, that controls the total logarithmic drop of the potential available before freezing. More efficient reheating shortens the kination era and reduces the distance traversed by the scalar field, whereas less efficient reheating allows a larger excursion before the field freezes. Once the field has frozen, its kinetic energy becomes negligible and its energy density is dominated by the potential,
\begin{align}
    \rho_\phi\simeq\tilde{V}(\phi_{\mathrm{freeze}})~,
\end{align}
so that the scalar field behaves as an effective cosmological constant until it thaws at late times.

The post-inflationary field excursion determines how much the potential can decrease between the onset of kination and the frozen configuration. We quantify this through the logarithmic potential drop:
\begin{align}
    D\equiv\left|\Delta\ln \tilde{V}\right|=\ln\left[\frac{\tilde{V}(\phi_{\mathrm{kin}})}{\tilde{V}(\phi_{\mathrm{freeze}})}\right].
\end{align}
Using the logarithmic slope introduced in Eq.~\eqref{eq:lambdaphi}, this may be written as
\begin{align}
    D=\int_{\phi_{\mathrm{kin}}}^{\phi_{\mathrm{freeze}}}\lambda_\phi(\phi)\,\frac{\dif\phi}{\m}~.
    \label{eq:potential-drop-integral}
\end{align}
It is useful to introduce the average logarithmic slope along the post-inflationary trajectory,
\begin{align}
    \bar{\lambda}\equiv\frac{1}{\Delta\phi}\int_{\phi_{\mathrm{kin}}}^{\phi_{\mathrm{freeze}}}\lambda_\phi(\phi)\,\dif\phi=D\frac{\m}{\Delta\phi}~,
    \label{eq:average-logarithmic-slope}
\end{align}
which provides a direct measure of the average steepness of the potential over the available field excursion. As can be seen, for a fixed energy difference between the inflationary and dark-energy scales, the required average slope is inversely proportional to the available field excursion. A shorter field excursion therefore demands a correspondingly steeper average slope, a requirement we confront with explicit realizations of $K(\phi)$ in the following sections.

%%%%%%%%%%%%%%%%%%%%%%%%%%%%%%%%%%%%%%%%%%%%%%%%%%%
%%%%%%%%%%%%%%%%%%%%%%%%%%%%%%%%%%%%%%%%%%%%%%%%%%%

%%%%%%%%%%%%%%%%%%%%%%%%%%%%%%%%%%%%%%%%
\section{\label{sec:single-exponential-model}Single-Exponential Model}
%%%%%%%%%%%%%%%%%%%%%%%%%%%%%%%%%%%%%%%%
\subsection{Definition and Potential}
Following Ref.~\cite{Park:2024ceu}, we consider an exponential nonminimal coupling:
\begin{align}
\label{eq:K_single_exp}
    K(\phi)\equiv \xi e^{-\phi/f_0}~,
\end{align}
for which
\begin{align}
    F_{,\phi}(\phi)=K_{,\phi}(\phi)=-\frac{K(\phi)}{f_0}~.
\end{align}
The positive dimensionless parameter $\xi$ can be absorbed by a constant shift of the scalar field. At the level of the full action we keep it explicit, since interaction terms contained in $\mathcal{L}_I$ need not respect the corresponding shift symmetry.\footnote{Possible effects associated with explicitly shift-symmetry-breaking interaction terms lie beyond the scope of the present analysis.}

The associated Einstein-frame potential is (see Eq.~\eqref{eq:thetildepotential-asymp})
\begin{align}
    \tilde{V}(\varphi)=V_0\left(1+\frac{e^{\phi(\varphi)/f_0}}{\xi}\right)^{-2}\simeq\begin{cases}
    V_0 \left(1-\frac{2}{\xi}e^{\phi/f_0}\right),
    & \phi/f_0 \ll \ln \xi~,\\[10pt]
    V_0 \xi^2 e^{-2\phi/f_0}~,
    & \phi/f_0 \gg \ln \xi~,
    \end{cases}
\end{align}
corresponding, respectively, to the nearly flat inflationary plateau and the runaway exponential potential that drives late-time quintessence. Throughout the background analysis presented in this work we fix
\begin{align}
    \xi=1~,
\end{align}
which simply sets the origin of the scalar-field coordinate without affecting either the inflationary dynamics or the subsequent post-inflationary evolution considered here. With this choice, negative field values correspond approximately to the inflationary plateau, while positive values interpolate toward the runaway quintessential regime. The kination epoch and the subsequent cosmological evolution therefore occur along the large-field branch.

%%%%%%%%%%%%%%% 
%%%%%%%%%%%%%%%

\subsection{\label{sec:inf-predictions-inflation}Inflationary Predictions}
We now examine the inflationary predictions of the model. Since inflation occurs on the plateau of the Einstein-frame potential, the dynamics are well described by the standard slow-roll approximation. In this section we derive the predictions for the scalar spectral index and tensor-to-scalar ratio as functions of the two parameters governing the inflationary dynamics, namely $f_0$ and the Einstein-frame number of $e$-folds $\tilde{N}_\star$ between horizon exit of the CMB pivot scale and the end of inflation.

The two slow-roll parameters associated with the Einstein-frame potential are defined as (see Ref.~\cite{Baumann:2009ds})
\begin{align}
    \epsilon_{\tilde{V}}
    &\equiv
    \frac{\m^2}{2}
    \left(
    \frac{\tilde{V}_{,\varphi}}{\tilde{V}}
    \right)^2~,\\
    \eta_{\tilde{V}}
    &\equiv
    \m^2\frac{\tilde{V}_{,\varphi\varphi}}{\tilde{V}}~.
\end{align}
Rather than working directly with the canonically normalized field $\varphi$, it is convenient to express these quantities in terms of the Jordan-frame field $\phi$ using
\begin{align}
    \frac{\mathrm{d}}{\mathrm{d}\varphi}=J^{-1}\frac{\mathrm{d}}{\mathrm{d}\phi}~,
\end{align}
where the field-space factor $J(\phi)=\mathrm{d}\varphi/\mathrm{d}\phi$ was introduced in Eq.~\eqref{eq:non-canonical-function}. For the exponential nonminimal coupling considered here,
\begin{align}
    J^2(\phi)=\frac{1}{1+e^{-\phi/f_0}}\left(1+\frac{3\m^2}{2f_0^2}\frac{e^{-2\phi/f_0}}{1+e^{-\phi/f_0}}\right).
\end{align}
Substituting this expression into the slow-roll definitions gives
\begin{align}
    \epsilon_{\tilde{V}}(\phi)
    &=
    \frac{2\m^2}{f_0^2}
    \left(
    1+e^{-\phi/f_0}
    +\frac{3\m^2}{2f_0^2}e^{-2\phi/f_0}
    \right)^{-1},
    \\[8pt]
    \nonumber
    \eta_{\tilde{V}}(\phi)
    &=
    \frac{\m^2}{f_0^2}
    \left[
    4+3e^{-\phi/f_0}
    -
    \left(1-\frac{3\m^2}{f_0^2}\right)e^{-2\phi/f_0}
    -
    \frac{3\m^2}{f_0^2}e^{-3\phi/f_0}
    \right]
    \\
    &\phantom{------------------}
    \times
    \left(
    1+e^{-\phi/f_0}
    +\frac{3\m^2}{2f_0^2}e^{-2\phi/f_0}
    \right)^{-2}.
\end{align}

\begin{figure}[t]
    \centering
    \includegraphics[width=1.00\textwidth]{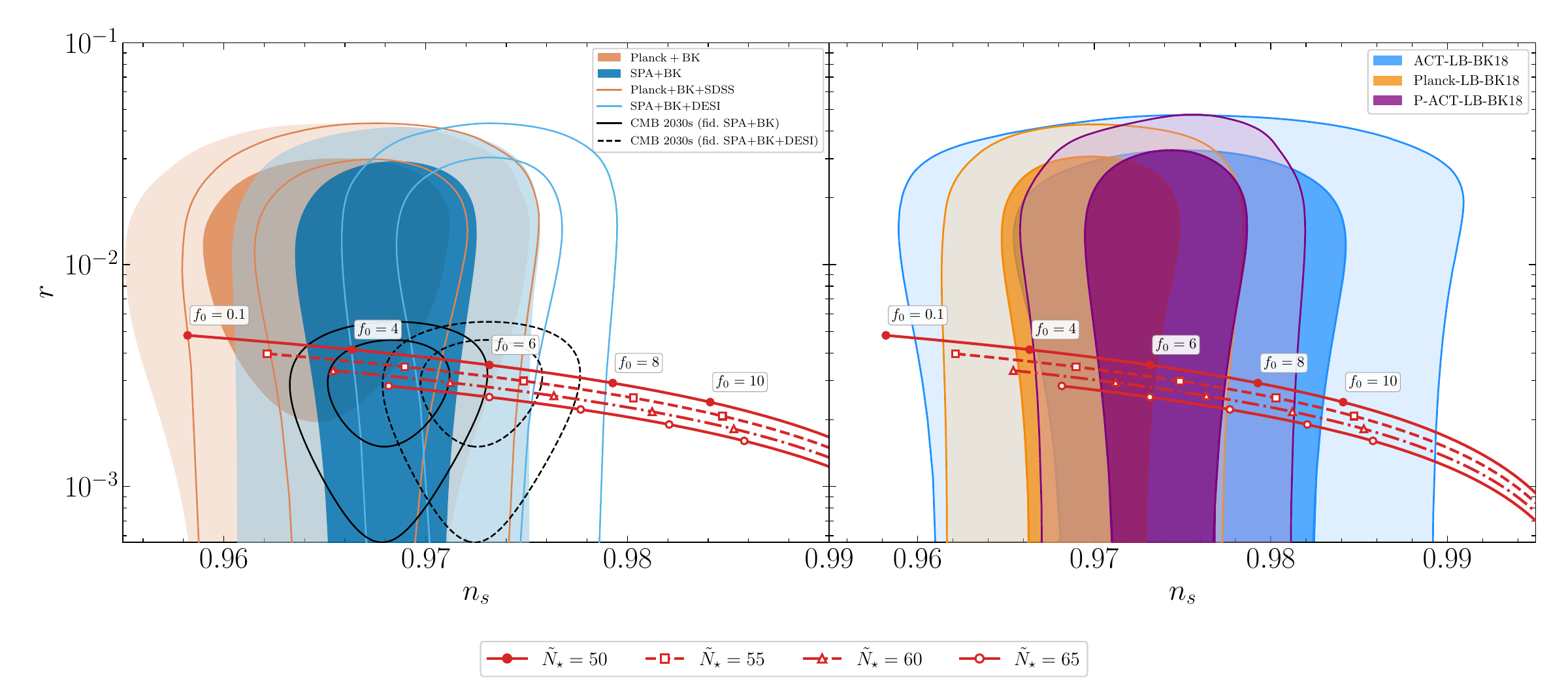}
    \caption{Inflationary predictions of the single-exponential model in the $n_s$--$\,r$ plane. The red curves correspond to different choices of the number of $e$-folds $\tilde N_\star$, traced as a function of $f_0$ (in Planck units), with markers indicating representative sub-Planckian and super-Planckian values of $f_0$. Shaded regions and contours denote the $68\%$ and $95\%$ confidence levels. The observational constraints were recovered from the vector paths of the published figures of Ref.~\cite{Balkenhol:2025wms} (left panel) and Ref.~\cite{AtacamaCosmologyTelescope:2025nti} (right panel), calibrated against the axis frames and tick positions. Both panels share the same logarithmic $r$ axis.}
    \label{fig:ns_vs_r_single_exp}
\end{figure}

The relation between the scalar-field value and the number of $e$-folds follows from the slow-roll equation:
\begin{align}
    \frac{\mathrm{d}\tilde{N}}{\mathrm{d}\phi}
    \simeq
    \frac{J^2}{\m^2}
    \frac{\tilde{V}}{\tilde{V}_{,\phi}}~,
\end{align}
which integrates to 
\begin{align}
    \tilde{N}(\phi)
    =
    \frac{f_0(\phi_{\mathrm{end}}-\phi)}{2\m^2}
    +
    \frac{3}{4}
    \left[
    e^{-\phi/f_0}
    -
    e^{-\phi_{\mathrm{end}}/f_0}
    +
    \ln
    \left(
    \frac{1+e^{-\phi_{\mathrm{end}}/f_0}}
    {1+e^{-\phi/f_0}}
    \right)
    \right],
\end{align}
with $\tilde{N}(\phi_{\mathrm{end}})=0$. Inflation occurs on the plateau of the potential, where $e^{-\phi/f_0} \gg 1$. In this regime the field value at horizon exit of the CMB pivot scale satisfies
\begin{align}
    \frac{\phi_{\star}}{f_0}
    \simeq
    \ln
    \left(
    \frac{3}{4\tilde{N}_{\star}}
    \right),
    \label{eq:pivot-scale-fvalue}
\end{align}
allowing the inflationary observables to be expressed directly in terms of the two parameters governing the inflationary dynamics: $f_0$ and $\tilde{N}_{\star}$. Using Eq.~\eqref{eq:pivot-scale-fvalue}, the slow-roll parameters become 
\begin{align}
    \epsilon_{\tilde{V}}(\tilde{N}_{\star})
    &\simeq
    \frac{3\m^2}{2f_0^2 \tilde{N}_{\star}}
    \left(
    1+\frac{2\m^2}{f_0^2}\tilde{N}_{\star}
    \right)^{-1}~,
    \\[8pt]
    \eta_{\tilde{V}}(\tilde{N}_{\star})
    &\simeq
    -
    \frac{\m^2}{f_0^2}
    \left(
    1+\frac{4\m^2}{f_0^2}\tilde{N}_{\star}
    \right)
    \left(
    1+\frac{2\m^2}{f_0^2}\tilde{N}_{\star}
    \right)^{-2}~,
\end{align}
from which the scalar spectral index $n_s$ and tensor-to-scalar ratio $r$ follow in the standard slow-roll approximation:
\begin{align}
    n_s-1
    &\simeq
    -6\epsilon_{\tilde{V}}
    +
    2\eta_{\tilde{V}}~,
    \\[8pt]
    r
    &\simeq
    16\epsilon_{\tilde{V}}~.
\end{align}
Hence,
\begin{align}
    n_s(\tilde{N}_{\star})-1
    &\simeq
    -
    \frac{\m^2}{f_0^2 \tilde{N}_{\star}}
    \left[
    9
    \left(
    1+\frac{2\m^2}{f_0^2}\tilde{N}_{\star}
    \right)
    +
    2\tilde{N}_{\star}
    \left(
    1+\frac{4\m^2}{f_0^2}\tilde{N}_{\star}
    \right)
    \right]
    \left(
    1+\frac{2\m^2}{f_0^2}\tilde{N}_{\star}
    \right)^{-2}~,
    \label{eq:ns-as-Nstartilde}
    \\
    r(\tilde{N}_{\star})
    &\simeq
    \frac{24\m^2}{f_0^2 \tilde{N}_{\star}}
    \left(
    1+\frac{2\m^2}{f_0^2}\tilde{N}_{\star}
    \right)^{-1}~.
\end{align}
At this stage, $f_0$ and $\tilde N_\star$ are treated as independent parameters. Equation~\eqref{eq:ns-as-Nstartilde} exhibits a feature that will prove relevant below. In the limit $f_0\to0$ the spectral index becomes independent of $f_0$,
\begin{align}
    n_s
    \to
    1-\frac{9+4\tilde{N}_\star}{2\tilde{N}_\star^{2}}~,
    \label{eq:ns-floor}
\end{align}
so that $n_s$ is bounded from below by a quantity fixed by $\tilde{N}_\star$ alone. This floor rises with $\tilde{N}_\star$, so that longer post-inflationary histories progressively exclude the low-$n_s$ region: for $\tilde{N}_\star=65$ the model cannot produce $n_s<0.968$ for any value of $f_0$. The model predictions in the $n_s$--$\,r$ plane are shown in Fig.~\ref{fig:ns_vs_r_single_exp}, together with the latest observational constraints from the CMB~\cite{Balkenhol:2025wms,Planck:2018jri,AtacamaCosmologyTelescope:2025nti}. We find that the model provides a good fit to current observations over a broad range of both sub-Planckian and super-Planckian values of $f_0$.

The observed amplitude of primordial scalar perturbations fixes the overall normalization of the potential. Within the slow-roll approximation \cite{Baumann:2009ds},
\begin{align}
    A_s(\phi)
    \simeq
    \frac{1}{24\pi^2\m^4}
    \frac{\tilde{V}(\phi)}
    {\epsilon_{\tilde{V}}(\phi)}~,
\end{align}
which, in the plateau limit, becomes
\begin{align}
    A_s(\tilde{N}_{\star})
    \simeq
    \frac{V_{0}}{36\pi^2\m^6}
    f_0^2
    \tilde{N}_{\star}
    \left(
    1+\frac{2\m^2}{f_0^2}\tilde{N}_{\star}
    \right).
\end{align}
Using the measured value of $A_s$,
we determine the corresponding normalization $V_0$,
with representative parameter choices listed in Table~\ref{tab:inflation_predictions}. Overall, the inflationary sector places only mild restrictions on the model parameters. The single-exponential construction naturally reproduces the observed values of $n_s$, $r$, and $A_s$ for a broad range of $f_0$. However, as we show below, the more restrictive constraints arise after inflation, where the reheating history determines the duration of kination and therefore the available scalar-field excursion connecting the inflationary plateau to the late-time quintessential regime. Anticipating that result, the matching condition derived in Sec.~\ref{sec:reh-match-sub} restricts
$f_0\lesssim0.3\,\m$, deep in the sub-Planckian regime, where the spectral index has saturated the bound of Eq.~\eqref{eq:ns-floor} and is controlled by $\tilde{N}_\star$ alone.

\begin{table}[ht]
\centering
\renewcommand{\arraystretch}{1.25}
\begin{tabular}{ccccc}
\hline\hline
$f_0/\m$ & $\tilde{N}_\star$ & $n_s$ & $r$ & $V_0/\m^4$ \\
\hline
  3.5 & 50 & 0.9647 & $4.28\times10^{-3}$ & $(1.329\pm0.019)\times10^{-10}$ \\
  4.5 & 50 & 0.9680 & $3.99\times10^{-3}$ & $(1.240\pm0.017)\times10^{-10}$ \\
  6.5 & 50 & 0.9748 & $3.37\times10^{-3}$ & $(1.049\pm0.015)\times10^{-10}$ \\
  7.5 & 50 & 0.9779 & $3.07\times10^{-3}$ & $(0.955\pm0.013)\times10^{-10}$ \\
\hline
  2.5 & 55 & 0.9651 & $3.75\times10^{-3}$ & $(1.166\pm0.016)\times10^{-10}$ \\
  4   & 55 & 0.9690 & $3.46\times10^{-3}$ & $(1.076\pm0.015)\times10^{-10}$ \\
  5.5 & 55 & 0.9734 & $3.11\times10^{-3}$ & $(0.967\pm0.014)\times10^{-10}$ \\
  7   & 55 & 0.9777 & $2.74\times10^{-3}$ & $(0.853\pm0.012)\times10^{-10}$ \\
\hline
  2   & 60 & 0.9671 & $3.23\times10^{-3}$ & $(1.002\pm0.014)\times10^{-10}$ \\
  3   & 60 & 0.9689 & $3.10\times10^{-3}$ & $(0.964\pm0.013)\times10^{-10}$ \\
  5.5 & 60 & 0.9751 & $2.66\times10^{-3}$ & $(0.827\pm0.012)\times10^{-10}$ \\
  7   & 60 & 0.9789 & $2.37\times10^{-3}$ & $(0.735\pm0.010)\times10^{-10}$ \\
\hline
  1.5 & 65 & 0.9690 & $2.79\times10^{-3}$ & $(0.868\pm0.012)\times10^{-10}$ \\
  2.5 & 65 & 0.9703 & $2.71\times10^{-3}$ & $(0.842\pm0.012)\times10^{-10}$ \\
  5   & 65 & 0.9754 & $2.38\times10^{-3}$ & $(0.740\pm0.010)\times10^{-10}$ \\
  6.5 & 65 & 0.9788 & $2.14\times10^{-3}$ & $(0.666\pm0.009)\times10^{-10}$ \\
\hline\hline
\end{tabular}
\caption{Slow-roll predictions for $n_s$, $r$, and the inflationary energy scale determined by $V_0$. For each $\tilde{N}_\star$, the first two values of $f_0$ lie in the lower part of the accessible $n_s$ range, favored by the SPA$+$BK combination of Fig.~\ref{fig:ns_vs_r_single_exp} (left panel), while the last two lie in the upper part, favored by the ACT DR6 combinations (right panel). $A_s$ is fixed from $\ln(10^{10}A_s)=3.044\pm0.014$ (TT,TE,EE+lowE+lensing) \cite{Planck:2018jri}, and the quoted uncertainty on $V_0$ propagates that of $A_s$.}
\label{tab:inflation_predictions}
\end{table}

%%%%%%%%%%%%%%% 
%%%%%%%%%%%%%%%

\subsection{Required Potential Drop}
A successful model of quintessential inflation must connect the inflationary energy scale to the observed dark-energy density through the post-inflationary evolution of the inflaton field. This immediately implies that the Einstein-frame potential must decrease by a large amount between the onset of kination and the epoch at which the field freezes. 

For a given value of $f_0$, the inflationary dynamics uniquely determine the potential energy at the onset of kination, $\tilde{V}(\phi_{\mathrm{kin}})$. Using the observed dark-energy density as a reference scale, we define the required logarithmic potential drop:
\begin{align}
    D_{\mathrm{req}}(f_0)
    \equiv
    \ln
    \left[
        \frac{
            \tilde V(\phi_{\mathrm{kin}};f_0)
        }{
            \rho_{\Lambda,0}
        }
    \right].
    \label{eq:required_potential_drop}
\end{align}
This quantity represents the total decrease that the potential must undergo in order to interpolate between the inflationary and dark-energy scales. It depends only on the inflationary normalization of the potential and is therefore determined before solving the subsequent cosmological evolution. The full numerical evolution performed later simply determines how closely the frozen scalar energy density reproduces the observed dark-energy abundance.

\begin{figure}[t]
    \centering
    \includegraphics[width=\linewidth]{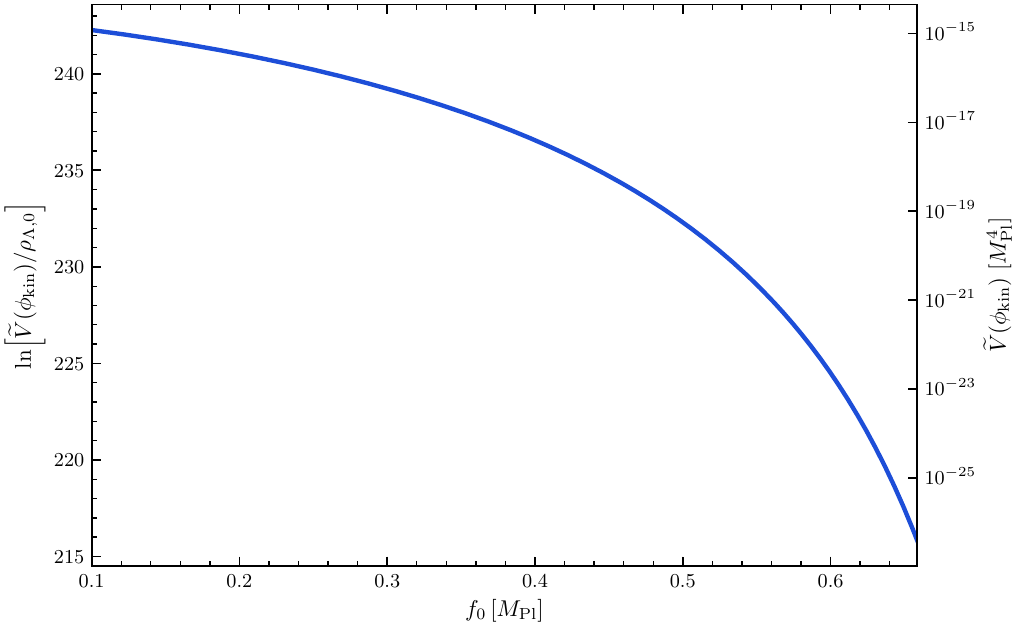}
    \caption{Potential energy at the onset of kination, $\tilde{V}(\phi_{\rm kin})$, and the logarithmic drop required to reach the present dark-energy density, \(D_{\rm req}=\ln[\tilde{V}(\phi_{\rm kin})/\rho_{\Lambda,0}]\), as functions of $f_0$ (in Planck units). The left vertical axis shows $D_{\rm req}$ and the right vertical axis the corresponding value of $\tilde{V}(\phi_{\rm kin})$ in units of $\m^4$. The two are related through \(\tilde{V}(\phi_{\rm kin})=\rho_{\Lambda,0}e^{D_{\rm req}}\), so the two axes represent the same curve.}
    \label{fig:Dvalues}
\end{figure}

Figure \ref{fig:Dvalues} shows the corresponding values of $\tilde{V}(\phi_{\mathrm{kin}})$ and $D_{\rm req}$ as functions of $f_0$. Over the range of parameters considered here, the potential at the onset of kination decreases from approximately $10^{-15}\m^4$ to $10^{-26}\m^4$, corresponding to a required logarithmic drop between $D_{\rm req}\simeq242$ and $D_{\rm req}\simeq216$. Even after inflation has reduced the vacuum energy by many orders of magnitude, the scalar potential must therefore continue to decrease by more than two hundred natural logarithmic units before reaching the present dark-energy scale. The post-inflationary evolution determines the field excursion available to realize this decrease. As discussed previously, the reheating history fixes the duration of kination and therefore the total displacement, $\Delta\phi$, between the onset of kination and the freezing point. Consequently, the required average logarithmic slope of the potential is
\begin{align}
    \bar{\lambda}_{\mathrm{req}}
    \equiv
    \frac{D_{\mathrm{req}}\,\m}{\Delta\phi}~,
    \label{eq:required-average-slope}
\end{align}
or, more explicitly, 
\begin{align}
    \bar{\lambda}_{\mathrm{req}}(f_0,T_{\mathrm{RH}})
    =
    \frac{D_{\mathrm{req}}(f_0)\,\m}
         {\Delta\phi(f_0,T_{\mathrm{RH}})}~.
    \label{eq:required-average-slope-f0-TRH}
\end{align}
The required potential drop is then fixed by inflation and the observed dark-energy scale, while the available field excursion is fixed by the reheating history. The reheating temperature therefore determines the minimum average steepness that the quintessence potential must possess after inflation.

%%%%%%%%%%%%%%% 
%%%%%%%%%%%%%%%

\subsection{\label{sec:reh-match-sub}Reheating and Matching}
The discussion above establishes two independent quantities. On the one hand, the inflationary energy scale together with the observed dark-energy density determines the required logarithmic potential drop, $D_{\rm req}$. On the other hand, the reheating history fixes the duration of kination and hence the available field excursion, $\Delta\phi$. A viable quintessential-inflation model must reconcile these two requirements through the shape of the scalar potential.

The actual average logarithmic slope of the potential along the post-inflationary trajectory is given by Eq.~\eqref{eq:average-logarithmic-slope}. Reproducing the observed dark-energy scale requires this average slope to coincide with the value dictated by the required potential drop:
\begin{align}
    \bar{\lambda}
    \simeq
    \bar{\lambda}_{\mathrm{req}}~,
    \label{eq:average-slope-matching-condition}
\end{align}
or, equivalently, 
\begin{align}
    \int_{\phi_{\mathrm{kin}}}^{\phi_{\mathrm{freeze}}}
    \lambda_\phi(\phi)\,\frac{\dif\phi}{\m}
    \simeq
    D_{\mathrm{req}}(f_0)~.
    \label{eq:potential-drop-matching-condition}
\end{align}
This matching condition provides a necessary consistency requirement for any model of quintessential inflation: the potential must be sufficiently steep, on average, to interpolate between the inflationary and dark-energy scales over the field excursion allowed by reheating. We first apply it to the minimal single-exponential model. Along the post-inflationary tail, the Einstein-frame potential is well approximated by
\begin{align}
    \tilde{V}(\phi)
    \simeq
    V_0 e^{-\lambda_0\phi/\m}~,
    \label{eq:quint-tail-singLe}
\end{align}
where
\begin{align}
    \lambda_0
    =
    \frac{2\m}{f_0}~.
    \label{eq:single-exp-slope}
\end{align}
Since the logarithmic slope is constant throughout the post-inflationary evolution,
\begin{align}
    \bar{\lambda}
    =
    \lambda_0
    =
    \frac{2\m}{f_0}~,
\end{align}
and the matching condition reduces to
\begin{align}
    \frac{2\m}{f_0}
    =
    \bar{\lambda}_{\mathrm{req}}(f_0,T_{\mathrm{RH}})
    =
    \frac{D_{\mathrm{req}}(f_0)\m}
         {\Delta\phi(f_0,T_{\mathrm{RH}})}~.
    \label{eq:single-exp-matching-condition}
\end{align}
Equation~\eqref{eq:single-exp-matching-condition} defines a one-dimensional locus in the $(f_0,T_{\mathrm{RH}})$ plane. For each value of $f_0$, only a particular reheating temperature produces the field excursion required to connect the inflationary plateau to the observed dark-energy scale. Solving it for the field excursion
gives $\Delta\phi=D_{\mathrm{req}}f_0/2$, so that the required duration of kination grows linearly with $f_0$. Together with the upper bound on
$\Delta\phi$ imposed by the GW constraint of Eq.~\eqref{eq:gw-upper-bound-Nkin}, this restricts the single-exponential model to $f_0\lesssim0.3\,\m$.

\begin{figure}[t]
    \centering
    \includegraphics[width=\linewidth]{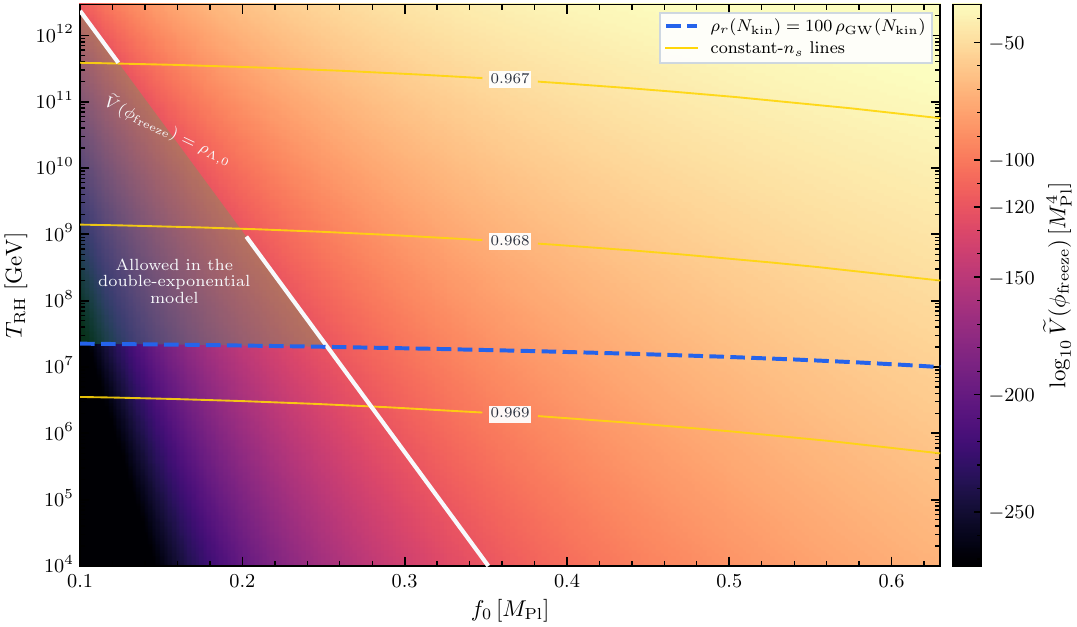}
    \caption{Reheating constraints and the frozen potential scale in the $f_0$--$T_{\rm RH}$ plane. The color map shows $\log_{10}[\tilde{V}(\phi_{\rm freeze})/\m^4]$ for the single-exponential model, and the yellow contours mark constant values of the scalar spectral index $n_s$. Along the white solid line the frozen potential matches the present dark-energy density, $\tilde{V}(\phi_{\rm freeze})=\rho_{\Lambda,0}$; to its left the single-exponential potential overshoots the required logarithmic drop, giving $\tilde{V}(\phi_{\rm freeze})<\rho_{\Lambda,0}$. This region can nonetheless be made viable in the double-exponential extension by adjusting the normalization $A$ of the shallow exponential contribution. The blue dashed line denotes the lower bound on $T_{\rm RH}$ obtained from the GW abundance condition, so that the shaded triangular region above it and to the left of the white line is viable in the double-exponential model.}
    \label{fig:reheating_constraints_f0_TRH}
\end{figure}

The resulting parameter space is shown in
Fig.~\ref{fig:reheating_constraints_f0_TRH}, where two constraints act simultaneously. The white curve, along which $\tilde{V}(\phi_{\mathrm{freeze}})=\rho_{\Lambda,0}$, identifies the combinations of $f_0$ and $T_{\mathrm{RH}}$ that reproduce the required total potential drop; to its left the potential overshoots and the frozen energy density falls below the observed dark-energy scale. The blue dashed curve is the GW bound of Eq.~\eqref{eq:gw-lower-bound-TRH}: below it the initial radiation abundance is too small, and the stochastic GW background exceeds the observational limit on $\N$. Only the region above the blue curve is therefore admissible, and within the single-exponential model only its intersection with the white curve reproduces the observed dark-energy density. As discussed below, the overshooting region can be made viable in the double-exponential extension by independently adjusting the late-time behavior of the potential.

The matching condition determines whether the potential can reproduce the required hierarchy between the inflationary and dark-energy scales; it does not, however, guarantee viable late-time cosmological dynamics. In particular, the same logarithmic slope that fixes the total potential drop also controls the asymptotic attractor of the scalar field. We now turn to this second requirement and show that it is here that the minimal single-exponential model encounters its fundamental difficulty.

%%%%%%%%%%%%%%% 
%%%%%%%%%%%%%%%

\subsection{Late-Time Attractors and the Failure of the Single-Exponential Model}
The previous section showed that the reheating temperature fixes the duration of kination and therefore the total post-inflationary field excursion. Matching the inflationary energy scale to the observed dark-energy density then determines the average logarithmic slope that the potential must exhibit along this trajectory. The remaining question is whether the corresponding single-exponential potential also gives rise to a viable late-time cosmology.

After kination, the scalar field rapidly loses kinetic energy and freezes during the radiation- and matter-dominated eras at the field value given in Eq.~\eqref{eq:sf_loses_kin_en}, with an approximately constant energy density
\begin{align}
    \rho_{\phi,\mathrm{freeze}}
    \simeq
    \tilde{V}(\phi_{\mathrm{freeze}})~.
\end{align}
This frozen configuration initially behaves as an effective cosmological constant. Once the background matter density has sufficiently diluted, however, the Hubble friction weakens and the field begins to evolve again. The subsequent dynamics are entirely determined by the asymptotic slope of the quintessential tail. For the single-exponential model, the potential takes the form in Eq.~\eqref{eq:quint-tail-singLe}, so the logarithmic slope is constant throughout the post-inflationary evolution. The cosmological dynamics of exponential quintessence are well understood from dynamical-systems analyses~\cite{Copeland:1997et,SavasArapoglu:2017pyh,Bahamonde:2017ize}. Depending on the value of $\lambda_0$, the scalar field approaches either a scalar-field-dominated solution or a background-scaling solution.

The scalar-field-dominated fixed point exists for 
\begin{align}
    \lambda_0^2<6~,
\end{align}
and is stable whenever
\begin{align}
    \lambda_0^2
    <
    3(1+w_b)~,
\end{align}
where $w_b$ is the EoS parameter of the dominant background fluid~\cite{Copeland:1997et,SavasArapoglu:2017pyh,Bahamonde:2017ize}. 
%\TT{[Add some explanation how these condition above are derived, or any refs.]}
The corresponding EoS is
\begin{align}
    w_\phi
    =
    -1
    +
    \frac{\lambda_0^2}{3}~,
    \qquad
    \Omega_\phi
    =
    1~,
\end{align}
yielding accelerated expansion only if
\begin{align}
    \lambda_0^2
    <
    2~.
\end{align}

Conversely, for
\begin{align}
    \lambda_0^2
    >
    3(1+w_b)~,
    \label{eq:scaling-existence-condition}
\end{align}
the attractor is instead the background-scaling solution:
\begin{align}
    w_\phi
    =
    w_b~,
    \qquad
    \Omega_\phi
    =
    \frac{3(1+w_b)}{\lambda_0^2}~.
    \label{eq:exponential-scaling-solution}
\end{align}
During matter domination this reduces to
\begin{align}
    w_\phi\to
    0~,
    \qquad
    \Omega_\phi\to
    \frac{3}{\lambda_0^2}~.
    \label{eq:matter-scaling-solution}
\end{align}
The crucial point is that, in the present construction, $\lambda_0$ is not an independent parameter. Once the reheating temperature has been fixed, the allowed field excursion is determined, and the matching condition of the previous section uniquely specifies the average logarithmic slope required to connect the inflationary plateau to the present dark-energy scale. In the single-exponential model, the average and local slopes coincide, $\bar \lambda = \lambda_0$, so the reheating history effectively fixes the late-time attractor of the model.

As anticipated in Sec.~\ref{sec:reh-match-sub}, the matching condition restricts $f_0\lesssim0.3\,\m$. Since $\lambda_0=2\m/f_0$, this implies
\begin{align}
    \lambda_0^2 \gtrsim 44 \gg 3~,
\end{align}
placing the scalar field unavoidably in the basin of the matter-scaling solution. Consequently, although the field initially behaves as thawing quintessence after emerging from its frozen state, this behavior is only transient. At late times, the evolution necessarily converges toward
\begin{align}
    w_\phi
    \to
    0~,
    \qquad
    \Omega_\phi
    \to
    \frac{3}{\lambda_0^2}
    =
    \frac{3f_0^2}{4\m^2}~,
\end{align}
rather than remaining as a dark-energy component.

\begin{figure}[t]
    \centering
    \includegraphics[width=0.95\linewidth]{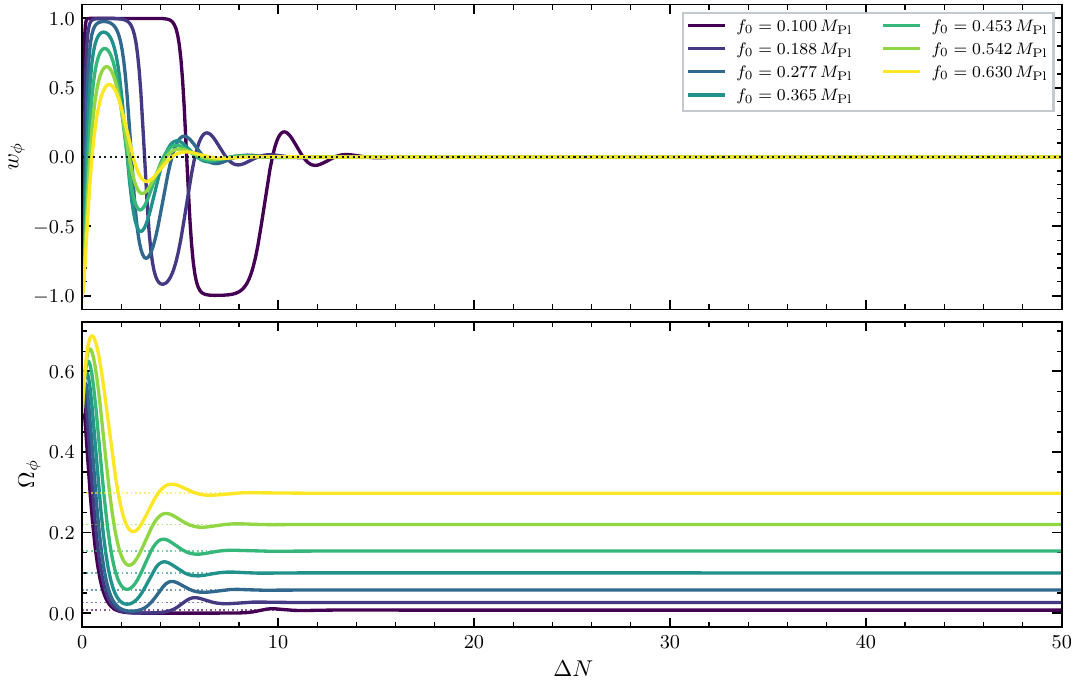}
    \caption{Late-time evolution in the single-exponential model for representative values of $f_0$ spanning, and extending somewhat beyond, the range allowed by the matching condition, $f_0\lesssim0.3\,\m$. All satisfy $\lambda_0^2>3$ and therefore approach the matter-scaling attractor. The upper panel shows the scalar-field EoS parameter $w_\phi$: the field is initially frozen with $w_\phi\simeq-1$ and, once it begins to roll, undergoes damped transient oscillations before settling on the matter-era scaling value $w_\phi=0$. The lower panel shows the scalar-field energy fraction $\Omega_\phi$, which asymptotically approaches $\Omega_\phi=3/\lambda_0^2=3f_0^2/(4\m^2)$; the dotted horizontal lines mark the corresponding analytic asymptotic values. Within the allowed range this gives $\Omega_\phi\lesssim0.07$, far below the observed dark-energy abundance. The horizontal coordinate is the shifted $e$-fold number $\Delta N\equiv N-N_{\rm ref}$, with $N_{\rm ref}$ the epoch at which $\rho_\phi=\rho_m$.}
\label{fig:single-exp-late-time-scaling}
\end{figure}

Figure \ref{fig:single-exp-late-time-scaling} illustrates this behavior for representative values of $f_0$. For ease of comparison, we introduce the shifted $e$-fold variable
\begin{align}
    \Delta N
    \equiv
    N-N_{\mathrm{ref}}~,
\end{align}
where $N_{\rm ref}$ denotes the epoch at which
\begin{align}
    \rho_\phi(N_{\mathrm{ref}})
    =
    \rho_m(N_{\mathrm{ref}})~.
\end{align}

The numerical solutions confirm the analytic picture. The scalar field remains frozen throughout most of the radiation- and matter-dominated eras, begins to thaw once its energy density becomes dynamically relevant, and subsequently undergoes damped oscillations before settling onto the matter-scaling attractor predicted by the dynamical-systems analysis.

The failure of the single-exponential model therefore does not simply reflect the well-known properties of exponential quintessence. Rather, it originates from the interplay between reheating and late-time cosmology. The GW bound fixes the minimum reheating temperature, which limits the duration of kination and therefore the available field excursion. This determines the average logarithmic slope required to bridge the inflationary and dark-energy scales. Because the single-exponential model identifies this average slope with the local slope governing the late-time dynamics, satisfying the matching condition inevitably drives the scalar field toward the matter-scaling attractor instead of a dark-energy-dominated solution. This observation also indicates how the obstruction can be removed. The required integrated potential drop must be preserved while reducing the local logarithmic slope near the present field value. In other words, one seeks (cf. Eq.~\eqref{eq:potential-drop-matching-condition})
\begin{align}
    \int_{\phi_{\mathrm{kin}}}^{\phi_{\mathrm{freeze}}}
    \lambda_\phi(\phi)\,\frac{\dif\phi}{\m}
    \simeq
    D_{\mathrm{req}}~,
\end{align}
together with
\begin{align}
    \lambda_\phi(\phi_{\mathrm{freeze}})
    \lesssim
    \mathcal O(1)~.
\end{align}

The required separation between the average and local logarithmic slopes may be achieved either by modifying the kinetic sector, thereby reducing the physical slope experienced by the canonically normalized field, or by deforming the potential itself. The former possibility is realized in kinetically stretched and $\alpha$-attractor models of quintessential inflation~\cite{Dimopoulos:2017zvq,Dimopoulos:2017tud,Brissenden:2023yko}, where the kinetic structure suppresses the effective late-time slope and enables thawing dark-energy behavior even for a steep underlying potential. In the present work, however, we adopt the latter approach and show that a minimal double-exponential deformation of the nonminimal coupling function naturally yields a steep average slope over the post-inflationary field excursion while keeping the asymptotic slope sufficiently shallow to support viable thawing quintessence.

%%%%%%%%%%%%%%%%%%%%%%%%%%%%%%%%%%%%%%%%%%%%%%%%%%%
%%%%%%%%%%%%%%%%%%%%%%%%%%%%%%%%%%%%%%%%%%%%%%%%%%%

%%%%%%%%%%%%%%%%%%%%%%%%%%%%%%%%%%%%%%%%
\section{\label{sec:double-exponential-extension}Double-Exponential Extension}
%%%%%%%%%%%%%%%%%%%%%%%%%%%%%%%%%%%%%%%%
\subsection{Definition and Decoupling of Sectors}
The single-exponential coupling of Sec.~\ref{sec:single-exponential-model} represents a special case where the runaway behavior of $K(\phi)$ is governed by a single scale. More generally, the coupling receives contributions from multiple scales and takes the form of a sum of exponentials with distinct slopes, of which the single-exponential ansatz retains only the steepest term. Such multi-exponential structures arise naturally for moduli in string and supergravity compactifications, where each contribution enters with a slope fixed by the underlying geometry~\cite{Copeland:2006wr}. Moreover, this same structure underlies double-exponential quintessence models that successfully interpolate between an early scaling regime and late-time cosmic acceleration~\cite{Barreiro:1999zs}. The minimal departure from the single-exponential model therefore retains two terms:
\begin{align}
    K(\phi)
    \equiv
    e^{-\phi/f_0}
    +
    A e^{-\phi/f_1}~,
    \qquad
    f_1>f_0~.
\end{align}
The second, shallower exponential becomes dynamically relevant only at large field values. Equivalently, defining the dimensionless slopes
\begin{align}
    \lambda_0
    \equiv
    \frac{2\m}{f_0}~,
    \qquad
    \lambda_1
    \equiv
    \frac{2\m}{f_1}~,
    \label{eq:theLambdas}
\end{align}
one has
\begin{align}
    \lambda_1
    <
    \lambda_0~,
    \label{eq:onebetterthantheotherone}
\end{align}
\emph{i.e.}, the second exponential is the shallower asymptotic runaway.

The additional exponential is introduced to modify the potential only near the field values relevant for late-time cosmology, while leaving the inflationary sector essentially unchanged. This requires
\begin{align}
    A e^{-\phi/f_1}
    \ll
    e^{-\phi/f_0}
\end{align}
throughout inflation and during most of the post-inflationary field excursion. In this regime,
\begin{align}
    K(\phi)
    \simeq
    e^{-\phi/f_0}~,
\end{align}
so that the inflationary dynamics, including the predictions for
$n_s$ and $r$, as well as most of the post-inflationary evolution,
coincide with those of the single-exponential model discussed in
Sec.~\ref{sec:single-exponential-model}. Since $f_1>f_0$, however, the second exponential decreases more slowly as the field evolves toward larger values. Consequently, it eventually becomes comparable to the first exponential near the freezing region, where it softens the logarithmic slope of the potential. This allows the average slope accumulated over the post-inflationary field excursion to remain sufficiently large to reproduce the required potential drop, while the local asymptotic slope becomes shallow enough to support viable thawing dark-energy evolution.

%%%%%%%%%%%%%%% 
%%%%%%%%%%%%%%%
\subsection{Late-Time Uplift, Freezing, and Thawing}
For the full Einstein-frame potential, the logarithmic slope is
\begin{align}
    \lambda_\phi
    =
    -\frac{2\m K_{,\phi}}{K(1+K)}~.
\end{align}
In the asymptotic regime, where $K\ll1$ and $\tilde V\simeq V_0K^2$, this becomes (see Eqs.~\eqref{eq:theLambdas} and \eqref{eq:onebetterthantheotherone})
\begin{align}
    \lambda_\phi(\phi)
    \simeq
    \frac{
        \lambda_0 e^{-\frac{\lambda_0\phi}{2\m}}
        +
        A\lambda_1 e^{-\frac{\lambda_1\phi}{2\m}}
    }{
        e^{-\frac{\lambda_0\phi}{2\m}}
        +
        A e^{-\frac{\lambda_1\phi}{2\m}}
    }~.
    \label{eq:double-exp-tail-slope}
\end{align}
Since $J\simeq1$ throughout the post-inflationary regime,
\begin{align}
    \lambda_\varphi
    \simeq
    \lambda_\phi~.
\end{align} 

Unlike the single-exponential model, the double-exponential potential contains an additional free parameter, $A$, which controls the field value at which the shallower exponential becomes important. We fix this parameter by requiring that the total logarithmic drop between the onset of kination and the freezing point reproduce the value required to connect the inflationary and dark-energy scales (see Eqs.~\eqref{eq:required_potential_drop} and \eqref{eq:potential-drop-matching-condition}). This condition provides an analytic estimate for $A$. In the full numerical evolution, we subsequently refine its value by requiring that the present-day expansion rate satisfy $H(T_0)=H_0$.

\begin{figure}[t]
    \centering
    \includegraphics[width=\linewidth]{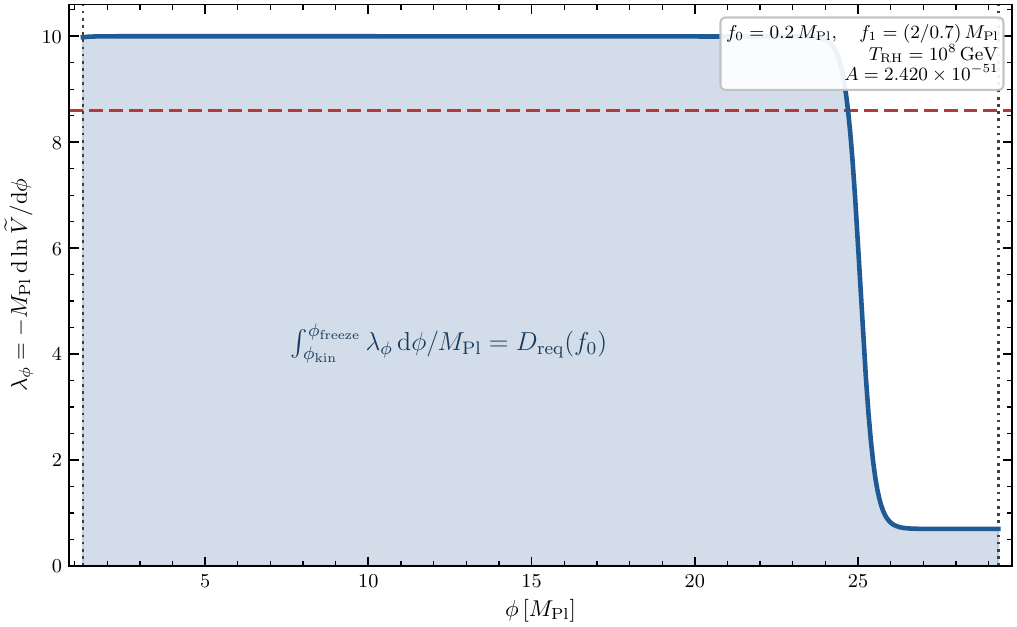}
    \caption{Logarithmic slope profile for the double-exponential potential. The slope remains large over most of the kination interval, preserving the required integrated area, and decreases near $\phi_{\mathrm{freeze}}$ toward the shallow late-time value $\lambda_1\simeq0.7$.}
    \label{fig:double_exp_lambda}
\end{figure}

The resulting logarithmic slope profiles are shown in Fig.~\ref{fig:double_exp_lambda}. The steep exponential dominates over most of the post-inflationary field excursion, thereby preserving the large integrated logarithmic slope required to connect the inflationary and dark-energy scales. Only near the freezing point does the shallower exponential become important, reducing the local logarithmic slope toward
\begin{align}
    \lambda_\phi(\phi_{\mathrm{freeze}})
    \simeq
    \lambda_1~.
\end{align}
For definiteness, we adopt
\begin{align}
    \lambda_1
    =
    0.7
\end{align}
as a benchmark late-time slope. This value is motivated by recent DESI-based analyses of exponential quintessence, which favor slopes of order $\lambda\simeq0.6$--$0.7$ \cite{Ramadan:2024kmn,Akrami:2025zlb}.

The double-exponential deformation therefore separates the quantity that determines the total potential drop from the quantity governing the late-time attractor; the average logarithmic slope remains fixed by the inflationary and dark-energy scales,
\begin{align}
    \bar{\lambda}
    &=
    \frac{1}{\Delta\phi}
    \int_{\phi_{\mathrm{kin}}}^{\phi_{\mathrm{freeze}}}
    \lambda_\phi(\phi)\,\dif\phi\simeq
    \frac{D_{\mathrm{req}}\,\m}{\Delta\phi}
    =
    \bar{\lambda}_{\mathrm{req}}~,
\end{align}
whereas the asymptotic slope is set to $\lambda_1=0.7$ by the second exponential. The matching condition and the late-time EoS are therefore controlled by two independent parameters, $f_0$ and $f_1$, so that a viable potential drop no longer forces the field onto the matter-scaling attractor that afflicts the single-exponential model.

%%%%%%%%%%%%%%% 
%%%%%%%%%%%%%%%

\subsection{Allowed Parameter Region}
The double-exponential extension enlarges the viable parameter space relative to the single-exponential model. The relevant constraints are summarized in Fig.~\ref{fig:reheating_constraints_f0_TRH}. As mentioned in the second-to-last paragraph in Sec.~\ref{sec:reh-match-sub}, the blue dashed line denotes the lower bound on the reheating temperature imposed by GW production. Parameter points below this line are excluded because the initial radiation abundance is insufficient to satisfy
\begin{align}
    \rho_{r,\mathrm{kin}}
    \geq
    100\,\rho_{{\mathrm{GW}},\mathrm{kin}}~.
\end{align}
The white solid line corresponds to the matching condition obtained in the single-exponential model, $\tilde{V}(\phi_{\mathrm{freeze}})=\rho_{\Lambda,0}$. To the left of this line, the steep exponential produces a logarithmic drop larger than required, yielding 
\begin{align}
    \tilde{V}(\phi_{\mathrm{freeze}})
    <
    \rho_{\Lambda,0}~,
\end{align}
so that these parameter combinations are excluded in the single-exponential scenario. The double-exponential deformation removes this restriction. By choosing an appropriate normalization $A$, the shallower exponential raises the potential only near the freezing point, reducing the total logarithmic drop while leaving the inflationary dynamics and the early post-inflationary evolution essentially unchanged. Consequently, every parameter point lying above the GW bound and to the left of the white matching line can be made viable by an appropriate choice of $A$. The shaded triangular region in Fig.~\ref{fig:reheating_constraints_f0_TRH} therefore represents the additional viable parameter space opened by the double-exponential deformation.

%%%%%%%%%%%%%%%%%%%%%%%%%%%%%%%%%%%%%%%%%%%%%%%%%%%
%%%%%%%%%%%%%%%%%%%%%%%%%%%%%%%%%%%%%%%%%%%%%%%%%%%

%%%%%%%%%%%%%%%%%%%%%%%%%%%%%%%%%%%%%%%%
\section{\label{sec:numerical-evolution}Numerical Evolution}
We now test the analytic picture presented above by numerically evolving the homogeneous cosmological background from the final stage of inflation to the present epoch. The complete Einstein-frame background equations, including radiation and pressureless matter, are summarized in Appendix~\ref{app:cov-field-eqs-background}.
%%%%%%%%%%%%%%%%%%%%%%%%%%%%%%%%%%%%%%%%
\subsection{\label{sec:back-initial-cond-section}Background Equations and Initial Conditions}
In the Einstein frame, the conformal transformation induces a direct coupling between the scalar field and nonrelativistic matter. The corresponding source term in the scalar-field equation is proportional to (see Eq.~\eqref{eq:varphi_2})
\begin{align}
    \frac{F_{,\varphi}}{2F}\tilde{\rho}_m~.
\end{align}
For the parameter range considered in this work, however, this contribution is negligible throughout the entire cosmological evolution. During the early Universe, the matter density is entirely subdominant, while at late times the field evolves in the asymptotic regime $K\ll1$, where $F\simeq1$ and $F_{,\varphi}/F$ is strongly suppressed. We therefore neglect this coupling and evolve the standard Einstein-frame equations,
\begin{align}
    \frac{\mathrm{d}^2 \varphi}{\mathrm{d}\tilde{\tau}^2}+3\tilde{H}\frac{\mathrm{d}\varphi}{\mathrm{d}\tilde{\tau}}+\tilde{V}_{,\varphi} =
    0~,
    \label{eq:numerical-KG}
\end{align}
together with 
\begin{align}
    &\frac{\mathrm{d}\tilde{\rho}_r}{\mathrm{d}\tilde{\tau}} +4\tilde{H} \tilde{\rho}_r = 0~,\\
    &\frac{\mathrm{d}\tilde{\rho}_m}{\mathrm{d}\tilde{\tau}}+3\tilde{H}\tilde{\rho}_{m} =0~,\\
    &3\tilde{H}^2 \m^2 = \frac{1}{2}\left(\frac{\mathrm{d}\varphi}{\mathrm{d}\tilde{\tau}}\right)^2 +\tilde{V}+\tilde{\rho}_r+\tilde{\rho}_m~,
    \label{eq:numerical-Friedmann}
\end{align}
where $\dif \tilde{\tau}$ denotes the conformally rescaled proper time of the so-called ``normal'' observer \cite{gourgoulhon:2012book}. The evolution of the radiation temperature includes the change in the effective relativistic degrees of freedom through entropy conservation,
\begin{align}
    g_{*s}(T)\tilde{a}^3T^3
    =
    {\mathrm{const.}}~.
\end{align}

The numerical integration begins during the final stage of slow-roll
inflation. During inflation and the transition to kination, the
Jordan- and Einstein-frame physical quantities can differ because the
conformal factor $F$ is still evolving, and the distinction between the
two frames must therefore be treated with care. The subsequent end of
inflation and the onset of kination are determined directly from the
Einstein-frame background evolution, with the latter defining the
reference epoch $\tilde{N}=0$. By this time, however,
\begin{align}
    K(\phi_{\rm kin})
    =
    \mathcal{O}(10^{-3})~,
\end{align}
so that $F(\phi_{\rm kin})=1+K(\phi_{\rm kin})\simeq1$. From the onset
of kination onward, the Einstein- and Jordan-frame time, Hubble
parameter, and energy densities therefore differ only by negligible
conformal corrections. At this reference epoch, we introduce a
subdominant radiation component according to
\begin{align}
    \tilde{\rho}_{r,\mathrm{kin}}
    =
    \Theta\tilde{\rho}_{\varphi,\mathrm{kin}}~,
    \label{eq:numerical-radiation-initial-condition}
\end{align}
where $\tilde{\rho}_{\varphi,\mathrm{kin}}$ is the Einstein-frame
scalar-field energy density at the onset of kination.

Rather than specifying the initial radiation fraction $\Theta$ directly, we treat the reheating temperature as the input parameter. Using Eq.~\eqref{eq:Tre}, the corresponding initial radiation fraction is
\begin{align}
\Theta
=
\left[
\frac{
\pi^2g_*(T_{\mathrm{RH}})
T_{\mathrm{RH}}^4
}{
30\tilde{\rho}_{\varphi,\mathrm{kin}}
}
\right]^{1/3}~.
\label{eq:numerical-theta-from-TRH}
\end{align}
This prescription guarantees that radiation overtakes the rapidly redshifting scalar kinetic energy at the specified reheating temperature.

The present matter density is fixed from the observed physical density parameter,
\begin{align}
    \rho_{m,0}
    = 3(\Omega_b h^2+\Omega_c h^2)H_{100}^2  \m^2~,
\end{align}
and evolved backward according to $\rho_m\propto a^{-3}$ to obtain the corresponding initial value at the onset of kination. $\Omega_b$ and $\Omega_c$ are the density parameters for baryon and dark matter, respectively, and $h \equiv H_0/H_{100}$ is the dimensionless Hubble parameter with
$H_{100}\equiv100~{\rm km}\,{\rm s}^{-1}{\rm Mpc}^{-1}$. As expected, the matter component remains completely negligible throughout inflation, kination, and radiation domination, becoming dynamically relevant only near the onset of the matter-dominated era.

%%%%%%%%%%%%%%% 
%%%%%%%%%%%%%%%

\subsection{Benchmark and Late-Time Thawing Evolution}
With the background equations and initial conditions in place, we turn to a representative benchmark of the double-exponential model. We consider
\begin{align}
    K(\phi)
    =
    e^{-\phi/f_0}
    +
    A e^{-\phi/f_1}~,
\end{align}
with
\begin{align}
    f_0
    &=
    0.2\,\m,
    \qquad
    \lambda_1
    =
    \{0.6,\,0.7,\,0.8\}~,
    \qquad
    f_1
    =
    \frac{2\m}{\lambda_1}~,
    \qquad
    T_{\mathrm{RH}}
    =
    10^8\,{\mathrm{GeV}}~.
    \label{eq:numerical-benchmark-parameters}
\end{align}
We adopt the present physical matter densities
\begin{align}
    \Omega_b h^2
    =
    0.02237~,
    \qquad
    \Omega_c h^2
    =
    0.1200~,
\end{align}
and define the present epoch by
\begin{align}
    T(\tilde{N}_0)
    =
    T_0
    =
    2.7255\,{\mathrm{K}}~.
    \label{eq:numerical-present-temperature}
\end{align}

\begin{figure}[p]
    \centering
    \includegraphics[width=0.95\linewidth]{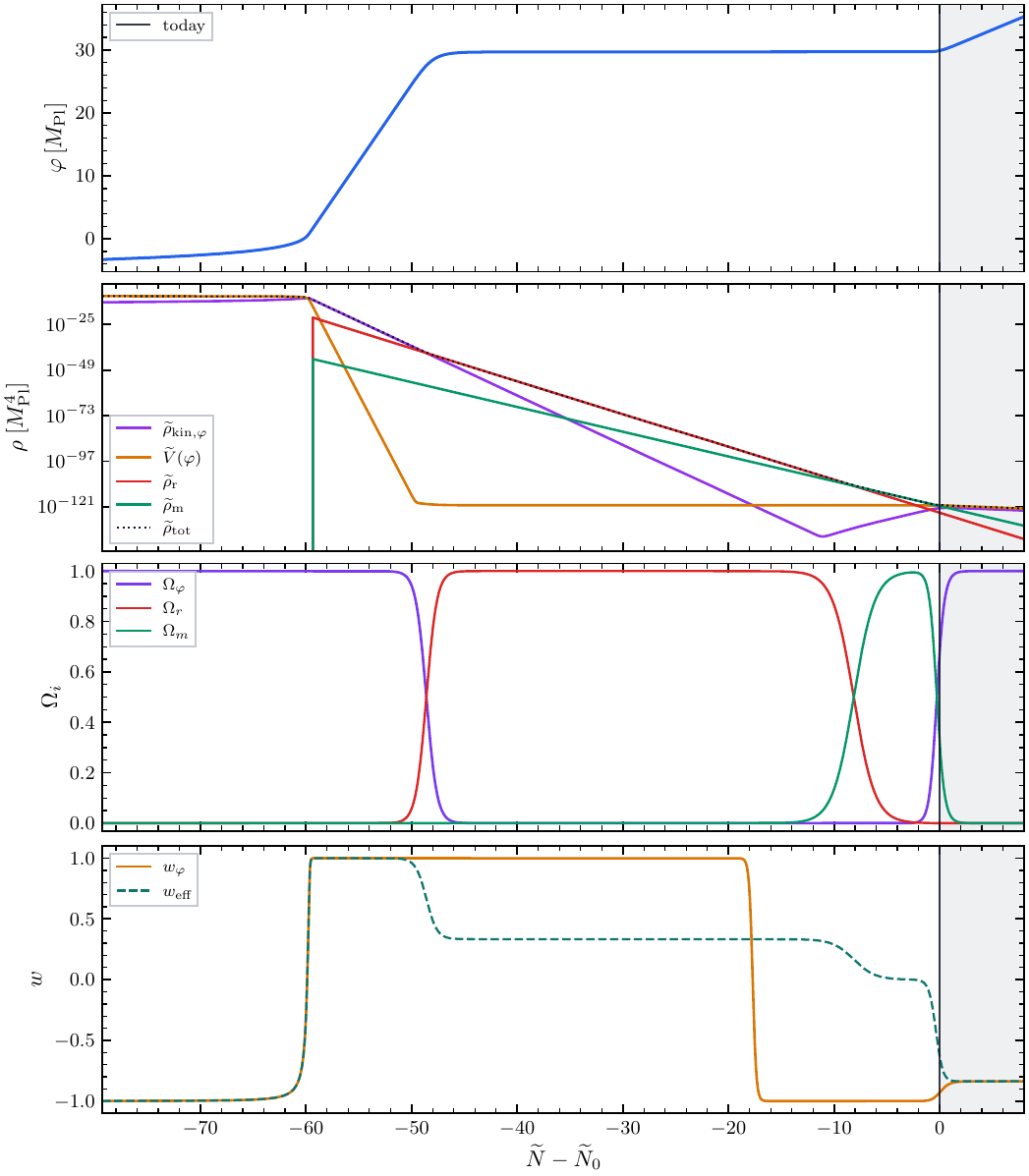}
    \caption{Numerical background evolution for the double-exponential benchmark of Eq.~\eqref{eq:numerical-benchmark-parameters} with $\lambda_1=0.7$, from the end of inflation to the present epoch. The panels show, respectively, the evolution of the scalar field, the energy densities and density fractions of the relevant components, and their EoS. The trajectory proceeds through kination, radiation domination, matter domination, and late-time scalar-field domination, the present epoch being fixed by $T=T_0=2.7255\,{\mathrm{K}}$.}
    \label{fig:background_evolution}
\end{figure}

For fixed $f_0$, $f_1$, and $T_{\mathrm{RH}}$, the normalization $A$ determines the relative importance of the shallow exponential and therefore fixes the late-time dark-energy scale. We determine $A$ by requiring that the Hubble parameter evaluated at the present CMB temperature reproduce the observed expansion rate,
\begin{align}
    \tilde{H}(\tilde{N}_0)
    \simeq
    67\,{\mathrm{km\,s^{-1}\,Mpc^{-1}}}~.
\end{align}
The corresponding values of $A$, for $\lambda_1=\{0.6,\,0.7,\,0.8\}$ respectively, are
\begin{align}
    A\simeq\left\{5.73\times10^{-52},\;
    2.51\times10^{-51},\;
    1.11\times10^{-50}\right\}~.
    \label{eq:numerical-benchmark-A}
\end{align}
Figure~\ref{fig:background_evolution} shows the full background
evolution for the central benchmark $\lambda_1=0.7$.
The panels show the scalar-field evolution, the energy densities and
density fractions of the relevant components, and the scalar and total
EoS, defined by
\begin{align}
    w_\varphi
    &\equiv
    \frac{
        (\mathrm{d}\varphi/\mathrm{d}\tilde{\tau})^2-2\tilde{V}
    }{
        (\mathrm{d}\varphi/\mathrm{d}\tilde{\tau})^2+2\tilde{V}
    }~,
    \\
    w_{\mathrm{eff}}
    &\equiv
    \frac{
        \tilde{p}_\varphi+\tilde{\rho}_r/3
    }{
        \tilde{\rho}_\varphi+\tilde{\rho}_r+\tilde{\rho}_m
    }
    =
    -1
    -
    \frac{2}{3}
    \frac{\dif \ln \tilde{H}}{\dif \tilde{N}}~.
    \label{eq:numerical-equations-of-state}
\end{align}
Following inflation, the scalar kinetic energy dominates the total energy density and the Universe enters a kination epoch with
\begin{align}
    w_{\mathrm{eff}}
    \simeq
    1~.
\end{align}
During kination, the scalar kinetic energy redshifts as $a^{-6}$, whereas radiation scales as $a^{-4}$. Radiation therefore inevitably overtakes the scalar component and becomes dominant at the prescribed reheating temperature. The subsequent radiation- and matter-dominated eras closely follow the standard thermal history. Throughout radiation domination and most of the matter era, Hubble friction efficiently suppresses the scalar motion. The field therefore remains nearly frozen, and its potential energy behaves as an effective cosmological constant. Only near the present epoch does the shallow exponential become dynamically relevant, allowing the field to thaw while preventing the evolution toward the matter-scaling attractor found in the single-exponential model.

\begin{figure}[t]
    \centering
    \includegraphics[width=\linewidth]{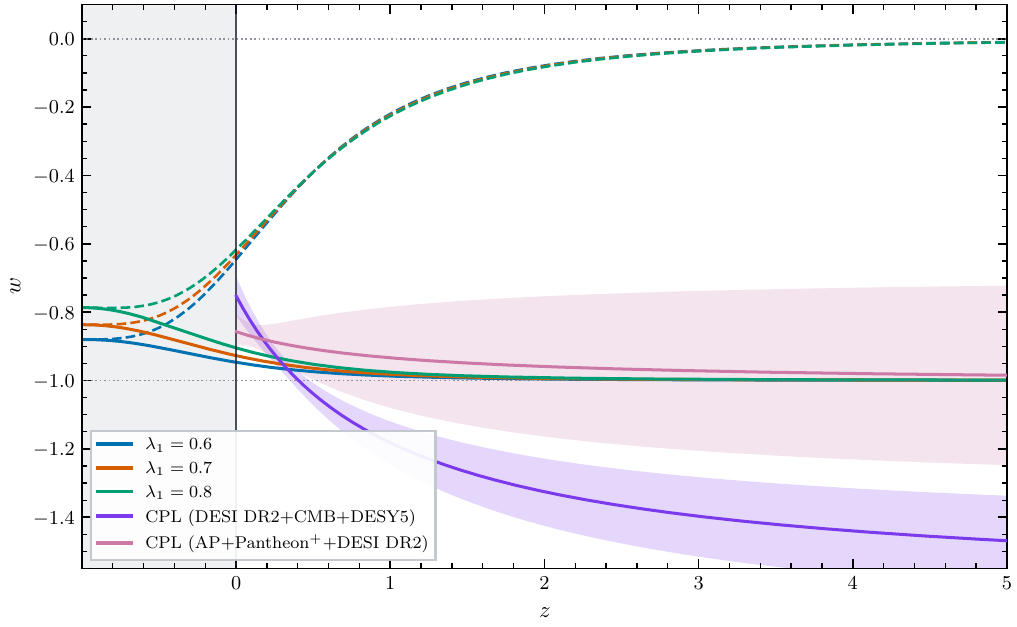}
    \caption{Late-time evolution of the scalar EoS parameter $w_\varphi(z)$ (solid) and the effective total EoS parameter $w_{\mathrm{eff}}(z)$ (dashed) for benchmark values $\lambda_1 = 0.6, 0.7$, and $0.8$. The scalar field remains nearly frozen with $w_\varphi \simeq -1$ until its energy density becomes comparable to that of matter, causing it to thaw at low redshift. For comparison, fits from flat CPL parameterizations are included from DESI DR2+Planck+DESY5~\cite{DESI:2025zgx} and AP+Pantheon+DESI DR2~\cite{Dong:2025ukl}, with shaded regions indicating $68\%$ confidence intervals. Because $w_\varphi$ is driven by a canonical scalar field, our model strictly satisfies $w_\varphi \ge -1$ at all times, whereas the DESI DR2+Planck+DESY5 CPL fit crosses the phantom divide at $z \simeq 0.4$. The CPL parameterizations are shown only for $z \ge 0$.}
    \label{fig:w_z}
\end{figure}

Figure~\ref{fig:w_z} compares the late-time evolution of $w_\varphi(z)$ and $w_{\mathrm{eff}}(z)$ for $\lambda_1=0.6$, $0.7$, and $0.8$. The redshift is related to the Einstein-frame number of $e$-folds $\tilde{N}$ by
\begin{align}
    1+z
    =
    e^{\tilde{N}_0-\tilde{N}}~.
\end{align}
Throughout radiation domination and almost the entire matter-dominated era, the scalar field remains frozen by Hubble friction. Consequently,
\begin{align}
    w_\varphi
    \simeq
    -1~,
\end{align}
and the scalar is effectively indistinguishable from a cosmological constant during this period. The field begins to evolve appreciably only after its potential energy becomes comparable to the matter density, at which point the Hubble damping weakens sufficiently for the scalar to roll along the shallow exponential tail. The thawing transition therefore occurs only at very low redshift, producing a correspondingly late departure from $w_\varphi=-1$. A larger value of $\lambda_1$ makes the asymptotic exponential steeper, leading to a slightly larger deviation of $w_\varphi$ from $-1$ at the present epoch, though the overall thawing behavior is preserved.

For the three benchmark values
$\lambda_1=\{0.6,\,0.7,\,0.8\}$,
the present-day values are
\begin{align}
    w_{\varphi,0}
    &\simeq \{-0.95,\,-0.93,\,-0.90\}, 
    \\
    w_{\mathrm{eff},0}
    &\simeq \{-0.65,\,-0.63,\,-0.62\}.
\end{align}

It is instructive to compare this behavior with the standard Chevallier–Polarski–Linder (CPL) parameterization \cite{Chevallier:2000qy,Linder:2002et},
\begin{align}
    w_{\mathrm{CPL}}(z) = w_0 + w_a \frac{z}{1+z}~,
\end{align}
which models dark energy as a smoothly evolving equation of state (EoS) over an extended redshift interval. For comparison, Fig.~\ref{fig:w_z} displays the best-fit flat CPL parameterization reported in Ref.~\cite{DESI:2025zgx},
\begin{align}
    w_0 = -0.752 \pm0.057~, \quad w_a = -0.86^{+0.23}_{-0.20}~, \quad \text{(DESI DR2+CMB+DESY5)}
\end{align}
alongside constraints from a recent tomographic Alcock–Paczyński (AP) test using redshift-space correlation functions \cite{Dong:2023jtk, Dong:2025ukl}:
\begin{align}
    w_0 = -0.857^{+0.051}_{-0.042}, \quad w_a = -0.153^{+0.347}_{-0.356}~. \quad \text{(AP+Pantheon$^{+}$+DESI DR2)}
\end{align}
The latter result implies no phantom crossing at redshifts $z < 0.7$.

In the DESI CPL fit, the strongly negative $w_a$ accommodates a present-day value $w_0 > -1$ while driving the EoS below the phantom divide ($w < -1$) at earlier redshifts ($z \gtrsim 0.4$). In contrast, the trajectory obtained in our model is qualitatively distinct. Because the scalar field $\varphi$ possesses a canonical kinetic term, its EoS strictly satisfies $w_\varphi \geq -1$ at all times and never crosses the phantom divide. Furthermore, with deviations from $w_\varphi = -1$ confined to low redshifts, the present value $w_{\varphi,0} \simeq -0.9$ remains closer to $-1$ than in the CPL-DESI fit.

The model thus realizes a classic thawing quintessence scenario: the scalar field behaves almost indistinguishably from a cosmological constant throughout most of cosmic history, departs from $w_\varphi = -1$ only near the present epoch, and maintains $w_{\mathrm{eff}} < -1/3$ to drive the observed cosmic acceleration.

%%%%%%%%%%%%%%%%%%%%%%%%%%%%%%%%%%%%%%%%%%%%%%%%%%%
%%%%%%%%%%%%%%%%%%%%%%%%%%%%%%%%%%%%%%%%%%%%%%%%%%%

%%%%%%%%%%%%%%%%%%%%%%%%%%%%%%%%%%%%%%%%
\section{\label{sec:conclusions}Conclusions}
%%%%%%%%%%%%%%%%%%%%%%%%%%%%%%%%%%%%%%%%
We have investigated quintessential inflation in a nonminimally coupled scalar-tensor theory, focusing on how the post-inflationary reheating history dictates and constrains the subsequent dark-energy evolution. Rather than committing to a specific reheating mechanism, we parameterized the radiation abundance at the onset of kination via its initial energy fraction, allowing a model-independent treatment of the post-inflationary background dynamics. In this framework, the reheating efficiency directly determines the duration of kination and, consequently, the total scalar-field excursion before the field freezes during radiation domination. Crucially, because the nonadiabatic transition from inflation to kination generates an irreducible stochastic gravitational-wave background, observational bounds on the effective number of relativistic species ($\Delta N_{\rm eff}$) impose a strict lower limit on the initial radiation abundance, and thus on the reheating temperature. This requirement translates directly into an upper bound on both the duration of kination and the available post-inflationary field excursion.

We then explored the implications of this bounded field excursion for the Einstein-frame potential. To bridge the vast energy hierarchy between the inflationary and present-day dark-energy scales within a limited field excursion, the potential must exhibit a sufficiently large average logarithmic slope. In the minimal single-exponential model, this average slope is identically equal to the asymptotic slope. While this successfully reproduces the required integrated potential drop, the resulting asymptotic slope is too steep to support viable late-time cosmic acceleration, causing the field to evolve toward a matter-scaling attractor instead. To resolve this tension, we introduced a minimal double-exponential deformation of the nonminimal coupling function, a structure naturally motivated by moduli dynamics in string and supergravity compactifications~\cite{Copeland:2006wr}. This construction decouples the average slope governing the total potential drop from the asymptotic slope controlling the late-time equation of state. The steeper exponential governs inflation and the post-inflationary kinetic drop, while the shallower exponential becomes dynamically relevant only near the freezing point, preserving the inflationary predictions while enabling a viable thawing quintessence regime at late times.

This analytical picture was confirmed through full numerical solutions of the background dynamics from the end of inflation to the present epoch. The numerical trajectories reproduce the required sequence of cosmological eras: kination, radiation domination, matter domination, and finally a thawing dark-energy phase. For the representative benchmark considered here, the present-day dark-energy equation of state is $w_{\varphi,0} \simeq -0.9$, a modest, observationally relevant departure from a cosmological constant that lies within the sensitivity of current and forthcoming dark-energy surveys.

Overall, our analysis demonstrates that the reheating history is not merely an auxiliary detail of early-Universe thermodynamics, but a fundamental arbiter of the viability of quintessential inflation. By regulating the duration of kination and the available scalar-field excursion, reheating constrains the global structure of the potential required to unify inflation with dark energy. Consequently, future improvements in constraints on the effective number of relativistic species, combined with increasingly precise determinations of the dark-energy equation of state, will provide powerful, complementary tests of this paradigm across cosmic history.

\vspace{10mm}

\noindent
\textbf{Acknowledgments}

\vspace{4mm}

\noindent
J.~J.~T.~D. thanks Samuel S\'anchez L\'opez for bringing Ref.~\cite{Balkenhol:2025wms} to his attention, which was used to compare against the latest constraints on the scalar spectral index and tensor-to-scalar ratio. S.~C.~P. and M.~G.~P. are supported by the National Research Foundation of Korea (NRF) grant funded by the Korea government (MSIT) (RS-2024-00340153). T.~T. is  supported in part by the JSPS KAKENHI Grant Number JP25K01004 and MEXT KAKENHI Grant Numbers JP25H01543 and JP26H00402. J.~J.~T.~D. acknowledges the financial support provided by FCT-Fundação para a Ciência e Tecnologia (FCT), I.P., through the Strategic Funding UID/04650/2025 and UID/04564/2025 and national funds with DOI identifiers 10.54499/2023.11681.PEX, 10.54499/2024.00249.CERN funded by measure RE-C06-i06.m02-``Reinforcement of funding for International Partnerships in Science, Technology and Innovation'' of the Recovery and Resilience Plan–RRP, within the framework of the financing contract signed between the Recover Portugal Mission Structure (EMRP) and the Foundation for Science and Technology I.P. (FCT), as an intermediate beneficiary, as well as the advanced computing projects 2024.00249.CERN.F1.

%%%%%%%%%%%%%%%%%%%%%%%%%%%%%%%%%%%%%%%%%%%%%%%%%%%
%%%%%%%%%%%%%%%%%%%%%%%%%%%%%%%%%%%%%%%%%%%%%%%%%%%

%%%%%%%%%%%%%%%%%%%%%%%%%%%%%%%%%%%%%%%%
\appendix
%%%%%%%%%%%%%%%%%%%%%%%%%%%%%%%%%%%%%%%%

%%%%%%%%%%%%%%%%%%%%%%%%%%%%%%%%%%%%%%%%%%%%%%%%%%%
%%%%%%%%%%%%%%%%%%%%%%%%%%%%%%%%%%%%%%%%%%%%%%%%%%%

%%%%%%%%%%%%%%%%%%%%%%%%%%%%%%%%%%%%%%%%
\section{\label{app:cov-field-eqs-background}Field Equations in Jordan and Einstein Frames}
In this appendix we derive the covariant field equations in the Jordan and Einstein frames and verify their equivalence under the Weyl rescaling of Sec.~\ref{sec:theoretical-framework}. We specialize them to a homogeneous and isotropic background employed in the numerical evolution of Sec.~\ref{sec:numerical-evolution}.
%%%%%%%%%%%%%%%%%%%%%%%%%%%%%%%%%%%%%%%%
\subsection{Jordan Frame}
Applying the principle of stationary action to $S[g_{\mu\nu},\phi,\Xi,\Psi] = S_{\textrm{grav}}[g_{\mu\nu},\phi]+S_{I}[g_{\mu\nu},\phi,\Xi]+S_{\textrm{M}}[g_{\mu\nu},\Psi]$ of Eq.~\eqref{eq:Jordan-frame-action-asitis}, varying with respect to the metric $g_{\mu\nu}$ and the scalar field $\phi$, and discarding boundary terms, we obtain
\begin{align}
    FG_{\mu\nu}-\left(\nabla_{\mu}\nabla_{\nu}-g_{\mu\nu}\Box\right)F&=\m^{-2}\left(\mathcal{T}_{\mu\nu}+T_{\mu\nu}^{\phi}\right),
    \label{eq:metric-fieldeqs-defframe}\\[8pt]
    \Box \phi -V_{,\phi}+\frac{\m^2}{2}F_{,\phi}R &= \mathcal{J}~,
    \label{eq:sf-equation-covariant}  
\end{align}
respectively, where $T_{\mu\nu}^{\phi}$ is the scalar field energy-momentum tensor (EMT):
\begin{equation}
    T_{\mu\nu}^{\phi} \equiv \partial_{\mu}\phi \partial_{\nu} \phi -g_{\mu\nu}\left[\frac{1}{2} (\partial\phi)^2+V\right],
    \label{eq:Defining-frame-field}
\end{equation}
and 
\begin{equation}
G_{\mu\nu} \equiv R_{\mu\nu}-\frac{1}{2} g_{\mu\nu}R
\end{equation}
is the Einstein tensor, which satisfies the contracted Bianchi identity $\nabla_{\nu} G^{\mu\nu}=0$ (see Ref.~\cite{Carroll:1997ar}), where $\nabla_{\mu}$ is the covariant derivative associated with the Levi-Civita connection of the Jordan-frame metric $g_{\mu\nu}$. The d'Alembertian is defined in Eq.~\eqref{eq:def-box-operator}, and 
\begin{align}
    \mathcal{T}_{\mu\nu} &\equiv g_{\mu\nu}\left(\mathcal{L}_I+\mathcal{L}_{\mathrm{M}}\right)-2\frac{\delta(\mathcal{L}_I+\mathcal{L}_{\mathrm{M}})}{\delta g^{\mu\nu}}~,
    \label{eq:def-stress-energy-tensor-interactions}\\
    \mathcal{J} &\equiv -\frac{\delta \mathcal{L}_I}{\delta \phi}~.
    \label{eq:source-term}
\end{align}
are, respectively, the matter EMT and the source term induced in the scalar-field equation by the interactions between $\phi$ and the matter sector. The former is conserved whenever $\mathcal{J}=0$ \cite{Koivisto:2005yk}: 
\begin{equation}
    \label{eq:exchange-matter-phi-dissipation-explicit}\nabla_{\mu}\tensor{\mathcal{T}}{^{\mu}_{\nu}} = -\mathcal{J} \partial_{\nu}\phi~.
\end{equation}
Under the assumption of local isotropy and homogeneity, one may adopt a perfect-fluid form for the EMT \cite{Maartens:1996vi}:
\begin{equation}
  \label{eq:perfect-fluid-form-emtensor}\mathcal{T}_{\mu\nu}\equiv \rho u_{\mu} u_{\nu}+ p h_{\mu\nu}~, 
\end{equation}
where $u_{\mu}$ is the normalized four-velocity field of the matter fluid, satisfying $u_{\mu} u^{\mu} = -1$, and $h_{\mu\nu}\equiv g_{\mu\nu}+u_{\mu}u_{\nu}$ is the projection tensor onto spatial hypersurfaces orthogonal to $u^{\mu}$. $\rho\equiv u^{\mu}u^{\nu}\mathcal{T}_{\mu\nu}$ is the energy density, and $p\equiv \frac{1}{3}h^{\mu\nu}\mathcal{T}_{\mu\nu}$ the (locally) isotropic pressure. Exploiting the properties of $u_{\mu}$ and $h_{\mu\nu}$, we arrive at the continuity equation:
\begin{align}
    u^{\mu}\partial_{\mu} \rho+\left(\nabla_{\mu}u^{\mu}\right)\left(\rho+p\right) = \mathcal{J}u^{\mu}\partial_{\mu} \phi~.
    \label{eq:continuity-matter}
\end{align}

We now assume a spatially flat Friedmann-Lema\^{i}tre-Robertson-Walker (FLRW) spacetime, with line element
\begin{equation}
    \label{eq:line-element-flrw-definingframe-anyframereally}
    \dif s^2 = -\mathcal{N}^2 \dif t^2 + a^2(t)\delta_{ij}\dif x^i \dif x^j~,
\end{equation}
where $\mathcal{N}$ is the lapse function, $t$ and $x^i$ denote the temporal and spatial coordinates, and $a(t)$ is the homogeneous scale factor. The metric, scalar-field, and continuity equations then read
\begin{align}
    \label{eq:f-eq-DF-Q}&3H^2\m^2 = \frac{1}{F}\left(\frac{\dot \phi^2}{2\mathcal{N}^2}+V-3H\m^2\frac{\dot F}{\mathcal{N}}+\rho\right),\\[8pt]
    \label{eq:tH-eq-DF-Q}&2\frac{\dot H}{\mathcal{N}}\m^2=-\frac{1}{F}\left[\frac{\dot \phi^2}{\mathcal{N}^2}-\m^2\left(H+\frac{\dot{\mathcal{N}}}{\mathcal{N}^2}\right)\frac{\dot F}{\mathcal{N}}+\m^2\frac{\ddot F}{\mathcal{N}^2}+\rho+p\right],\\[8pt]
    \label{eq:sf-eq-DF-Q}&\frac{\ddot \phi}{\mathcal{N}^2}+\left(3H-\frac{\dot{\mathcal{N}}}{\mathcal{N}^2}\right)\frac{\dot \phi}{\mathcal{N}}+V_{,\phi}-3\m^2F_{,\phi}\left(\frac{\dot H}{\mathcal{N}}+2H^2\right) = -\mathcal{J}~,\\[8pt]
    \label{eq:rad-eq-DF-Q}&\frac{\dot{\rho}}{\mathcal{N}} =-3H\left(\rho+p\right) +\mathcal{J}\frac{\dot \phi}{\mathcal{N}}~,
\end{align}
respectively. Here $\phi$ is understood to be homogeneous, as are all field-dependent quantities, including the energy density, pressure, and the source $\mathcal{J}$. Following the definitions of $\rho$ and $p$ given above Eq.~\eqref{eq:continuity-matter}, the energy density and pressure of the scalar field $\phi$ are
\begin{align}
    \label{eq:rhoofphi}\rho_{\phi}&=\frac{\dot \phi^2}{2\mathcal{N}^2}+V~,\\
    \label{eq:presofphi}p_{\phi}&=\frac{\dot \phi^2}{2\mathcal{N}^2}-V~.
\end{align}

We may specialize the matter sector to a mixture of radiation and pressureless matter:
\begin{align}
    \label{eq:split-energy-momentum-tensor-radiation-matter}\mathcal{T}_{\mu\nu}
    =
    \mathcal{T}^{r}_{\mu\nu}
    +
    \mathcal{T}^{m}_{\mu\nu}~,
    \ \ \mathrm{where} \ \
    p_r=\frac{1}{3}\rho_r~,
    \ \ \mathrm{and} \ \
    p_m=0~.
\end{align}
If we take the subset $\Xi$ of matter fields to constitute the pressureless component, while radiation belongs to the minimally coupled sector $\mathcal{L}_{\mathrm{M}}$, the continuity equation~\eqref{eq:rad-eq-DF-Q} splits into
\begin{align}
    \frac{\dot{\rho}_r}{\mathcal{N}}&=
    -4H\rho_r~,
    \label{eq:rad-continuity-DF}
    \\[8pt]
    \frac{\dot{\rho}_m}{\mathcal{N}}&=
    -3H\rho_m+\mathcal{J}\frac{\dot{\phi}}{\mathcal{N}}~,
    \label{eq:matter-continuity-DF}
\end{align}
where $\rho=\rho_r+\rho_m$ and $p=p_r+p_m$. In the absence of a direct coupling, $\mathcal{J}\to0$, both components are independently conserved and redshift as $\rho_r\propto a^{-4}$ and $\rho_m\propto a^{-3}$, respectively. This is the regime relevant to the numerical evolution of Sec.~\ref{sec:numerical-evolution}, where the residual scalar-matter coupling is negligible throughout the cosmological history.

%%%%%%%%%%%%%%% 
%%%%%%%%%%%%%%%

\subsection{Einstein Frame}
Reformulating the gravitational action through the Weyl rescaling of Eq.~\eqref{eq:Weyl-rescaling-metrictensor}, we obtain from the resulting action, Eq.~\eqref{eq:Einstein-frame-action}, the field equations (recall that $F=F(\phi)$):
\begin{align}
    \label{eq:metric-rescaled-Einstein}&\m^2\tilde{G}_{\mu\nu} = \tilde{\mathcal{T}}_{\mu\nu}+\tilde{T}^{\phi}_{\mu\nu}~,\\[8pt]
    \label{eq:scalar-rescaled-Einstein}&\tilde{\Box}\phi -F\tilde{V}_{,\phi}+\frac{F_{,\phi}}{2}\left(\tilde{\rho}-3\tilde{p}\right)+\frac{3\m^2}{2}F_{,\phi}\tilde{\Box}\ln F -\frac{F_{,\phi}}{2F}(\tilde{\partial} \phi)^2=\frac{\mathcal{J}}{F}~,
\end{align}
where
\begin{align}
   \label{eq:tilde-frame-metric-field-eqs}\tilde{G}_{\mu\nu}&\equiv \tilde{R}_{\mu\nu}-\frac{1}{2}\tilde{g}_{\mu\nu}\tilde{R}~,\\[8pt]
   \tilde{T}^{\phi}_{\mu\nu} &\equiv \frac{1}{F}T^{\phi}_{\mu\nu}+\frac{3\m^2}{2}\left[\partial_{\mu}\ln F\partial_{\nu}\ln F-\frac{1}{2}\tilde{g}_{\mu\nu}(\tilde{\partial} \ln F)^2\right],\\[8pt]
   \tilde{\mathcal{T}}_{\mu\nu} &\equiv \tilde{g}_{\mu\nu}\left(\tilde{\mathcal{L}}_I+\tilde{\mathcal{L}}_{\mathrm{M}}\right)-2\frac{\delta(\tilde{\mathcal{L}}_I+\tilde{\mathcal{L}}_{\mathrm{M}})}{\delta \tilde{g}^{\mu\nu}}~,
\end{align}
and
\begin{align}
    \tilde{\Box} \equiv \tilde{g}^{\mu\nu}\tilde{\nabla}_{\mu}\tilde{\nabla}_{\nu}~,
\end{align}
with $\tilde{\nabla}_{\mu}$ being the covariant derivative associated with the Levi-Civita connection of the Einstein-frame metric $\tilde{g}_{\mu\nu}$, and the notation $(\tilde{\partial} \ln F)^2$ was established in Eq.~\eqref{eq:notation-tilde-square-kinetic}. Also, $\nabla_{\mu} \phi = \partial_{\mu} \phi = \tilde{\nabla}_{\mu} \phi$ (and similarly for any scalar function). We define $\tilde{\rho}\equiv \tilde{u}^{\mu}\tilde{u}^{\nu}\tilde{\mathcal{T}}_{\mu\nu}$ and $\tilde{p}\equiv \frac{1}{3}\tilde{h}^{\mu\nu}\tilde{\mathcal{T}}_{\mu\nu}$, where $\tilde{u}_{\mu}$ is the normalized four-velocity field of the matter fluid satisfying\footnote{It is precisely this normalization condition that makes $u^{\mu}$ frame dependent.} $\tilde{u}_{\mu} \tilde{u}^{\mu} = -1$, with $\tilde{u}_{\mu}\equiv \tilde{g}_{\mu\nu}\tilde{u}^{\nu}$, and $\tilde{h}_{\mu\nu}\equiv \tilde{g}_{\mu\nu}+\tilde{u}_{\mu}\tilde{u}_{\nu}$ projects onto spatial hypersurfaces orthogonal to $\tilde{u}^{\mu}$. Because the Einstein tensor is identically divergenceless with respect to its associated Levi-Civita connection \cite{Carroll:1997ar}, from Eq.~\eqref{eq:metric-rescaled-Einstein} we obtain the continuity equation (also using the properties of $\tilde{u}^{\mu}$ and $\tilde{h}_{\mu\nu}$)
\begin{align}
    \label{eq:continuity-rescaled-Einstein}\tilde{u}^{\mu}\partial_{\mu}\tilde{\rho}+(\tilde{\nabla}_{\mu}\tilde{u}^{\mu})\left(\tilde{\rho}+\tilde{p}\right)+\frac{1}{2}\left(\tilde{\rho}-3\tilde{p}\right)\tilde{u}^{\mu}\partial_{\mu}\ln F = \frac{\mathcal{J}}{F^2}\tilde{u}^{\mu}\partial_{\mu}\phi~.
\end{align}

Alternatively, we can arrive at Eqs.~\eqref{eq:metric-rescaled-Einstein}, \eqref{eq:scalar-rescaled-Einstein}, and \eqref{eq:continuity-rescaled-Einstein} through the Weyl rescaling of the respective Jordan-frame field equations \eqref{eq:metric-fieldeqs-defframe}, \eqref{eq:sf-equation-covariant}, and \eqref{eq:continuity-matter}, noting that \cite{Dabrowski:2008kx}
\begin{align}
    &\tilde{G}_{\mu\nu} = G_{\mu\nu}-\nabla_{\mu}\partial_{\nu}\ln F +\frac{1}{2}\partial_{\mu}\ln F \partial_{\nu}\ln F +\frac{1}{2}g_{\mu\nu}\left[2\Box \ln F +\frac{1}{2}(\partial\ln F)^2\right],\\[8pt]
    \label{eq:rescaling-energymomentumtensor}&\tilde{\mathcal{T}}_{\mu\nu}\equiv \frac{1}{F}\mathcal{T}_{\mu\nu}~,\\[8pt]
    &\tilde{\rho}=F^{-2}\rho~,
    \label{eq:rescale-rhom}\\[8pt]
    &\tilde{p}=F^{-2}p~,
    \label{eq:rescale-pm}\\[8pt]
    &\tilde{u}^{\mu} = F^{-1/2}u^{\mu}~,\\[8pt]
    &\tilde{\nabla}_{\mu}\tilde{u}^{\mu} = \frac{1}{\sqrt{F}}\left(\nabla_{\mu}u^{\mu}+\frac{3}{2}u^{\mu}\nabla_{\mu}\ln F\right),
\end{align}
and the transformation of the Ricci scalar in Eq.~\eqref{eq:Ricci-rescaling-Weyl}. Additionally, in order to obtain Eq.~\eqref{eq:scalar-rescaled-Einstein}, one needs to eliminate $\tilde{R}$ in the scalar-field equation by means of the trace of Eq.~\eqref{eq:metric-rescaled-Einstein}. Through the scalar-field redefinition in Eq.~\eqref{eq:field-redefinition-canonical}, one arrives at the field equations
\begin{align}
    \m^2 \tilde{G}_{\mu\nu} &= \tilde{\mathcal{T}}_{\mu\nu}+\tilde{T}^{\varphi}_{\mu\nu}~,
    \label{eq:METRIC-eq-rescaledandredefined}\\[8pt]
    \tilde{\Box}\varphi -\tilde{V}_{,\varphi} &=\frac{1}{F}\left[\frac{\mathcal{J}}{FJ}-\frac{F_{,\varphi}}{2}\left(\tilde{\rho}-3\tilde{p}\right)\right],
    \label{eq:sf-eq-rescaledandredefined}\\[8pt]
    \tilde{u}^{\mu}\partial_{\mu}\tilde{\rho}+(\tilde{\nabla}_{\mu}\tilde{u}^{\mu})\left(\tilde{\rho}+\tilde{p}\right) &= \frac{1}{F}\left[\frac{\mathcal{J}}{FJ}-\frac{F_{,\varphi}}{2}\left(\tilde{\rho}-3\tilde{p}\right)\right]\tilde{u}^{\mu}\partial_{\mu}\varphi~,
    \label{eq:continuity-eq-rescaledandredefined}
\end{align}
where
\begin{align}
    \label{eq:redefined-scalar-field-EMT}\tilde{T}^{\varphi}_{\mu\nu} \equiv \partial_{\mu}\varphi \partial_{\nu}\varphi -\tilde{g}_{\mu\nu}\left[\frac{1}{2}(\tilde{\partial}\varphi)^2 +\tilde{V}\right],
\end{align}
is the redefined scalar-field EMT, which takes its canonical form. 

The Weyl rescaling in Eq.~\eqref{eq:Weyl-rescaling-metrictensor} transforms the FLRW line element (cf. Eq.~\eqref{eq:line-element-flrw-definingframe-anyframereally}) into 
\begin{align}
    \label{eq:line-element-flrw-Einsteinframe-actually}\dif \tilde{s}^2 = -\tilde{\mathcal{N}}^2\dif t^2+\tilde{a}^2(t)\delta_{ij}\dif x^i\dif x^j~.
\end{align}
The lapse function and the scale factor are related by\footnote{Using the rescaling of the scale factor, the number of $e$-folds can be shown to transform as 
\begin{align}
    \tilde{N}\equiv \ln\left(\frac{\tilde{a}}{\tilde{a}_0}\right) = N+\frac{1}{2}\ln\left(\frac{F}{F_0}\right).
\end{align}}
\begin{align}
    \tilde{\mathcal{N}}&=\sqrt{F}\mathcal{N}~,
    \label{eq:transformation-background-lapse-conformal} \\[8pt]
    \tilde{a} &= \sqrt{F}a~.
    \label{eq:transformation-background-scale-factor}
\end{align}
The spacetime slicing is kept unchanged, meaning that the coordinate time $t$ is not redefined. The proper time of a normal observer $\tau$ (\emph{i.e.} an observer whose four-velocity coincides with the vector field that is orthogonal to the slicing \cite{gourgoulhon:2012book}) does transform according to \cite{Karciauskas:2022jzd}  
\begin{align}
    \dif\tilde{\tau} = \tilde{\mathcal{N}} \dif t = \sqrt{F}\dif\tau~.
    \label{hattau-def}
\end{align} 
The Einstein-frame Hubble parameter is related to its defining-frame counterpart via the relation \cite{Chiba:2013mha,Kuusk:2016rso,Karciauskas:2022jzd}
\begin{align}
    \label{eq:conformal-transformation-Hubbleexprate}\tilde{H} = \frac{\dot{\tilde{a}}}{\tilde{\mathcal{N}}\tilde{a}} = \frac{H}{\sqrt{F}}\left(1+\frac{\dot{F}}{2H\mathcal{N}F}\right).
\end{align}
Here too we consider homogeneous quantities, despite retaining the same notation as for the spacetime-dependent ones (for example, the nonminimal coupling $F$). Under these transformations (and Eqs.~\eqref{eq:rescale-rhom} and \eqref{eq:rescale-pm}) the system of equations \eqref{eq:f-eq-DF-Q}--\eqref{eq:rad-eq-DF-Q} takes the following form:
\begin{align}
    \label{eq:Friedmann-Einstein-frame}
    &3\m^2\tilde{H}^2= \frac{1}{2}\left(\frac{\dif\varphi}{\dif\tilde{\tau}}\right)^2+\tilde{V}(\varphi)+\tilde{\rho}~,\\[8pt]
    &2\frac{\dif\tilde{H}}{\dif\tilde{\tau}}\m^2 = -\left[\left(\frac{\dif\varphi}{\dif\tilde{\tau}}\right)^2+\tilde{\rho}+\tilde{p}\right],\label{eq:Hdot_df} \\[8pt]
    &\frac{\dif^2\varphi}{\dif\tilde{\tau}^2}+3\tilde{H}\frac{\dif\varphi}{\dif\tilde{\tau}}+\tilde{V}_{,\varphi}(\varphi)=-\frac{1}{F}\left[\frac{\mathcal{J}}{FJ}-\frac{F_{,\varphi}}{2}\left(\tilde{\rho}-3\tilde{p}\right)\right],\label{eq:varphi_2}\\[8pt]
    \label{eq:rho_r_2}&\frac{\dif\tilde{\rho}}{\dif\tilde{\tau}}=-3\tilde{H}\left(\tilde{\rho}+\tilde{p}\right)+\frac{1}{F}\left[\frac{\mathcal{J}}{FJ}-\frac{F_{,\varphi}}{2}\left(\tilde{\rho}-3\tilde{p}\right)\right]\frac{\dif \varphi}{\dif \tilde{\tau}}~.
\end{align}
These may also be obtained from the covariant field
equations~\eqref{eq:METRIC-eq-rescaledandredefined}--\eqref{eq:continuity-eq-rescaledandredefined} by inserting the FLRW metric of Eq.~\eqref{eq:line-element-flrw-Einsteinframe-actually}. Because radiation has a traceless energy-momentum tensor, $\tilde\rho_r-3\tilde p_r=0$ (see Eq.~\eqref{eq:split-energy-momentum-tensor-radiation-matter}), its continuity equation is unaffected by the conformal coupling and retains the standard form. Eqs.~\eqref{eq:rad-continuity-DF} and \eqref{eq:matter-continuity-DF} then become 
\begin{align}
    \frac{\dif\tilde{\rho}_r}{\dif\tilde{\tau}}&=
    -4\tilde{H}\tilde{\rho}_r~,
    \label{eq:rad-continuity-EF}
    \\
    \frac{\dif\tilde{\rho}_m}{\dif\tilde{\tau}}&=
    -3\tilde{H}\tilde{\rho}_m+\frac{1}{F}\left(\frac{\mathcal{J}}{FJ}-\frac{F_{,\varphi}}{2}\tilde{\rho}_m\right)\frac{\dif \varphi}{\dif \tilde{\tau}}~.
    \label{eq:matter-continuity-EF}
\end{align}
Even if $\mathcal{J}=0$, the pressureless component in the rescaled frame does not redshift as $\tilde{a}^{-3}$ due to the presence of the conformal coupling. In the model considered in this work, however, the nonminimal-coupling contribution is negligible, as explained in Sec.~\ref{sec:back-initial-cond-section}. 

Finally, by analogy with the definitions of $\tilde{\rho}$ and $\tilde{p}$ and using the EMT of Eq.~\eqref{eq:redefined-scalar-field-EMT}, we obtain the energy density and pressure of $\varphi$ (cf.~Eqs.~\eqref{eq:rhoofphi} and \eqref{eq:presofphi}):
\begin{align}
    \label{eq:rhoofvphi}\tilde{\rho}_{\varphi}&=\frac{1}{2}\left(\frac{\dif \varphi}{\dif \tilde{\tau}}\right)^2+\tilde{V}~,\\[8pt]
    \label{eq:presofvphi}\tilde{p}_{\varphi}&=\frac{1}{2}\left(\frac{\dif \varphi}{\dif \tilde{\tau}}\right)^2-\tilde{V}~.
\end{align}

%%%%%%%%%%%%%%%%%%%%%%%%%%%%%%%%%%%%%%%%%%%%%%%%%%%
%%%%%%%%%%%%%%%%%%%%%%%%%%%%%%%%%%%%%%%%%%%%%%%%%%%

%\clearpage 
\bibliography{ref}

\end{document}